\documentclass[]{jfm}

\usepackage{graphicx,caption}
\usepackage{newtxtext}
\usepackage{newtxmath}
\usepackage{natbib}
\usepackage{hyperref}
\usepackage{textcomp}
\usepackage{amsmath,mathrsfs}
\hypersetup{
    colorlinks = true,
    urlcolor   = blue,
    citecolor  = black,
}

\newcommand{\RomanNumeralCaps}[1]
\linenumbers

\newcounter{roman_braket}  

\newenvironment{my_indent}
  {\begin{list}
    {\roman{roman_braket}}
    {\usecounter{roman_braket}
     \setlength{\labelwidth}{2em}
     \setlength{\labelsep}{0.1em}
     \setlength{\itemsep}{2pt}
     \setlength{\leftmargin}{1.2cm}
     \setlength{\rightmargin}{0.8cm}
     \setlength{\itemindent}{0.1em} 
     \setlength{\topsep}{2pt}
     
    }
  }
{\end{list}}

\newcommand{\eps}{\varepsilon}

\newcommand{\bx}{{\boldsymbol{x}}}
\newcommand{\bu}{{\boldsymbol{u}}}
\newcommand{\bv}{{\boldsymbol{v}}}
\newcommand{\beq}{\begin{equation}}
\newcommand{\eeq}{\end{equation}}
\newcommand{\delij}{\updelta}
\newcommand{\opL}{\mathcal{L}}
\newcommand{\R}{\mathcal{R}}
\newcommand{\A}{{\mathcal A}}
\newcommand{\N}{\mathcal{N}}
\newcommand{\NL}{\Gamma}
\newcommand {\Ov}{\Upsilon}
\newcommand{\bq}{\boldsymbol{q}}
\newcommand{\bhq}{\boldsymbol{\hat{q}}}
\DeclareMathOperator*{\argmax}{arg\,max}

\title{Non-normal weakly nonlinear analysis: asymptotic consistency and non-universality}

\author{Matthew McCormack\aff{1} \corresp{\email{matthew.mccormack@ed.ac.uk}}, Gregory P. Chini \aff{2} \& Rich R. Kerswell\aff{3} }

\affiliation{
\aff{1}School of Mathematics and Maxwell Institute for Mathematical Sciences, University of Edinburgh, UK, 
\aff{2}Department of Mechanical Engineering and Program in Integrated Applied Mathematics, University of New Hampshire, Durham, NH 03824 USA
\aff{3}Department of Applied Mathematics and Theoretical Physics, University of Cambridge, Wilberforce Road, Cambridge CB3 0WA, UK}

\begin{document}
\maketitle

\begin{abstract}

Non-normality can induce large transient growth in linearly stable systems. Determining whether this growth triggers a transition in the underlying nonlinear system, however, requires understanding the interaction between non-normality and nonlinearity.
Here, we develop a weakly nonlinear theory for linearly-stable, non-normal systems subject to harmonic forcing, enabling a systematic analysis of this interaction.
Following Ducimeti\`ere et al. (\emph{J. Fluid Mech.}, vol. 947, 2022, A43), we define a formal small parameter $\varepsilon$ as the reciprocal of the system's maximum linear amplification.  However, we ensure asymptotic consistency by providing a framework that naturally adapts to the underlying structure of the system.  The approach is applied to a harmonically forced channel flow and to a two-dimensional model mimicking the structure of the Orr--Sommerfeld--Squire equations.  Unlike classical weakly nonlinear analysis near bifurcation points, the resulting amplitude equations are non-universal.  In fact, a single linear mode amplified by the non-normality can nonlinearly excite a multi-modal and multi-frequency response at leading-order, which is system- or even regime-specific.  Nevertheless, the method yields asymptotically consistent amplitude equations that capture this complexity provided a limit in which $\varepsilon\rightarrow0$ can be identified.
As the forcing amplitude increases, the reduced equations capture stable nonlinear states emerging from the laminar flow, their subsequent bifurcations, and their eventual collision with the boundary of their basin of attraction.  Thus, the amplitude equations can capture subcritical transitions driven by forcing and varied initial conditions and enable the identification of critical parameters beyond which no stable weakly nonlinear state exists.

\end{abstract}


%
%
\newpage
\section{Introduction}

The use of amplitude equations to study dynamical systems near bifurcation points has been highly successful and is now well-established textbook material (e.g. \cite{Iooss80, Guckenheimer83, Crawford91, Cross93}). In fluid mechanics, the approach can trace its roots back to the development of so-called ``weakly nonlinear analysis'' (WNA) in the late 1950s \citep{Gorkov57, Malkus_Veronis_1958, Stuart1958non, Stuart1960non, watson1960non}. The essence of this approach is to capture the dynamics in the neighbourhood of a bifurcation point via a low-order asymptotic expansion. The key ingredient is that the amplitude of the bifurcating eigenfunction vanishes at the bifurcation point and so by continuity is small in its neighbourhood, making a small-amplitude expansion viable. This is usually only pursued until the first effect of nonlinearity has been captured and the character of the bifurcation is clear (i.e. whether it is supercritical or subcritical). The approach works well for the Navier--Stokes equations since: (i) the nonlinearity is only quadratic in the velocity field (as opposed to being, say, a transcendental function) so only a small number of nonlinearly generated fields are typically involved; and (ii) the `small' parameter $\eps$ of the expansion only enters through the amplitude and adjusted control parameter of the problem (e.g. $Re-Re_c=\mathcal{O}(\eps^2)$ for a symmetry-breaking bifurcation at the critical Reynolds number $Re_c$, where $Re$ is the Reynolds number), so all the operators to be inverted are independent of $\eps$ to leading order. Consequently, the amplitude equations derived are universal for a given type of  bifurcation, and then the focus of the analysis is on computing the sign of the numerical coefficient of the nonlinear amplitude term.

Given the success of WNA, it has been natural to try to generalise the approach to different settings where an appropriately defined amplitude is small.  One plausible variation is to assume that the bifurcation occurs at `infinity' in $Re$ and to look for an amplitude expansion where the amplitude tends to zero as $Re \rightarrow \infty$ (e.g. \cite{Smith82} in Newtonian pipe flow), or even to pursue a high-order amplitude expansion looking for convergence (e.g. viscoelastic pipe flow \citep{Meulenbroek04} and most recently viscoelastic channel flow \citep{Morozov19}) regardless whether the  amplitude decays or not as $Re \rightarrow \infty$.

Most recently, \cite{ducimetiere2022weak} have derived amplitude equations for weakly-forced, linearly-stable flow states that nevertheless are prone to large but transient growth of disturbances; i.e. the linear operator governing disturbance evolution around the base flow state is highly non-normal. Here, the reciprocal of the peak amplification factor of these disturbances is used to define a small parameter, and then classical WNA-style analysis is pushed through. Initially, \cite{Ducimetiere_stochastic_2022} and \cite{Ducimetiere_Lamb-Oseen_2023} artificially singularised the linear operator to mirror what happens at a bifurcation, but a subsequent reformulation built around a more appropriate singular value decomposition of the linear operator \citep{Ducimetiere_PRE_2025} avoids this. Interestingly, they appear to achieve a good match with numerical data (e.g. figure 5 in \cite{ducimetiere2022weak}) even though, as shown herein, their amplitude equations are not strictly rationally derived nor consistently applied to the data (see \S\ref{sec:duc_compare}, \ref{sec:PPF_pre_asy} and \ref{sec:comparison_model_duc}). 
The underlying motivation for our study is to revisit the work of \cite{ducimetiere2022weak} and \cite{Ducimetiere_PRE_2025} to understand this puzzling behaviour and, more significantly, to develop an asymptotically consistent formalism for the weakly nonlinear analysis of strongly non-normal dynamical systems.

Given that non-normality has long been appreciated to be generic in shear flows (or more accurately that the Navier--Stokes operator linearised around a steady  shear flow typically is non-normal: \cite{Kel-87, Orr-07}; and more recently e.g. \cite{Farrell_orr, Butler92, Trefethen93, Reddy93, Schmid07}), it is perhaps surprising that the theory for deriving the relevant amplitude equations has not already been laid out. In what follows, however, we will see that the theory is much more intricate than classical WNA and does not obviously lead to a universal amplitude equation structure owing to three additional complications. The first is that the asymptotic behaviour of the chosen small parameter (the smallest singular value of the linearised operator or equivalently the inverse of the largest singular value of the associated resolvent) may not be clear either in the asymptotic limit $Re \rightarrow \infty$ or across forcing characteristics (e.g. the forcing frequency $\omega$).
Secondly, the leading singular value of the resolvent may not be asymptotically separated from the remaining singular values, implying that multiple singular vector fields can interact at leading order, which typically leads to a system of nonlinearly coupled equations for each leading-order amplitude.  Finally, and perhaps causing the most difficulty, inverting various operators 
and carrying out the projections needed to derive the amplitude equations can each introduce non-trivial dependence on $\eps$. Inner products of different singular vectors, for example, can scale with some power of $\eps$ rather than being generically $\mathcal{O}(1)$ or precisely zero by some symmetry considerations, as in WNA. It is this ingredient that contributes most to making the analysis case-dependent.

Despite these complications, the fact that non-normal behaviour is ubiquitous, e.g. also arising in plasma physics \citep{Camporeale_plasma,Friedman_plasma_2014,Landreman_Plunk_Dorland_2015,Friedman_plasma_2015}, magnetohydrodynamics
\citep{KRASNOV_ZIKANOV_ROSSI_BOECK_2010,Squire_MRI,MacTaggart_2018,fraser2026nonmodal},
neural dynamics \citep{di2018non}, species dynamics \citep{nicoletti2018non}, ecology \citep{townley2007predicting}, economics \citep{sornette2023non}, and broadly across the dynamics of networks \citep{asllani2018structure,muolo2019patterns,Nicoletti_2019}, indicates, at least to us, that this weakly nonlinear theory is still worth pursuing.

The plan of the paper is as follows. In \S\ref{sec:gen_system} we outline a general approach for deriving asymptotically consistent amplitude equations for strongly non-normal systems.  This approach is then applied to a channel (plane Poiseuille) flow in \S\ref{sec:PPF}, which has previously been investigated by \citet{ducimetiere2022weak}, allowing us to directly compare the methodologies.  The complications in deriving these amplitude equations are then examined in more detail  by treating a much simpler system of two ordinary differential equations (ODEs) in \S\ref{sec:model_prb}, where all the information needed to reduce the system asymptotically can be obtained analytically.  Our conclusions are given in \S\ref{sec:conclusion}.

%
%

\section{Asymptotic reduction of a general system}\label{sec:gen_system}

We start by considering a general, harmonically-forced, nonlinear dynamical system
\begin{equation}
    \partial_t \bq = \opL \bq + \N(\bq,\bq) + \delta({\boldsymbol{\hat{f}}}e^{\mathrm{i}\omega t} + {c.c.})
    \label{eq:eq_main}
\end{equation}
where $\opL$ is a linear operator, $\N$ is a purely quadratically nonlinear operator as in the Navier--Stokes equations (so $\N( {\bf 0},{\bf 0} )={\bf 0} $ allowing $\bq={\bf 0}$ to be an unforced solution, e.g. corresponding to a steady laminar flow), $\omega$ is the frequency of the forcing and 
\beq
\| \boldsymbol{\hat{f}} \|_2 := \,\sqrt{ \int |\boldsymbol{\hat{f}} |^2 \,\mathrm{d}^3\bx }\,=\,1,
\eeq
so the amplitude of the forcing is set by $\delta$ (\,${c.c.}$ denotes complex conjugate). 
The key objective here is developing a weakly nonlinear picture of the response $\bq$ to this forcing when $\opL$ is a `highly' non-normal (to be defined below in (\ref{highly})) but asymptotically stable operator, meaning that all its eigenvalues have negative real parts. 

%
%
\subsection{Linear response} \label{sec:linear}

In the absence of the nonlinear term $\N$, the response is solely at the forcing frequency. Noting that there are no eigenvalues of $\opL$ on the imaginary axis (by assumption), and writing
$\bq = \bhq e^{\mathrm{i}\omega t} + {c.c.}$, this linear forced response is
\begin{equation}
    \bhq = \R(\mathrm{i}\omega)\delta\boldsymbol{\hat{f}}:=(\mathrm{i}\omega I - \opL)^{-1}\delta\boldsymbol{\hat{f}},
    \label{eq:resolvent}
\end{equation}
where $\mathcal{R}(i\omega)$ is the resolvent operator at a frequency $\omega$ ($I$ is the identity).  A magnified  linear response ensues when the resolvent norm
\beq
\|\mathcal{R}(\mathrm{i}\omega)\|:= \sup_{{\bf x}\neq {\bf 0}} \frac{\|\mathcal{R}(\mathrm{i} \omega) {\bf x}\|_2}{\| {\bf x}\|_2}
\eeq
is large. This can occur if $i \omega$ is close to an eigenvalue (a resonance) for general $\opL$ or even away from any eigenvalues if $\opL$ is non-normal \citep[e.g.][]{trefethen2020spectra}, i.e.~if
\beq 
\opL \opL^T \neq \opL^T \opL
\eeq
where $\opL^T$ is the adjoint of $\opL$ under some inner product. The case of interest here is the latter where the eigenvalues are an $\mathcal{O}(\eps^0)$ distance  away from $i \omega$ (and remain to the left of the imaginary axis) yet there is an asymptotic limit in which
\beq 
\eps := \frac{1}{\|\mathcal{R}(\mathrm{i}\omega(\eps))\|} \rightarrow 0. 
\label{highly}
\eeq
 For example, $\eps$ could  be some inverse power of the Reynolds number $Re$ in the Navier--Stokes equations, with $Re \rightarrow \infty$ and $\omega_*(Re)$ the forcing frequency that elicits the largest linear response at a given $Re$; that is  
\beq 
\omega_*(Re) =\argmax_\omega \|\mathcal{R}(\mathrm{i}\omega;Re)\|.
\eeq
Generally, there is a non-trivial forcing frequency set $\{\omega(0)\}$ that achieves (\ref{highly}). The simplest situation is to choose a fixed $\omega$ in this set for $\eps \geq 0$ and then consider the limit $\eps \rightarrow 0$; this approach is pursued below.

The challenge is then to understand: (i) how to build a weakly nonlinear theory for the response $\bq$ as $\eps \rightarrow 0$, {\em and} (ii) for what forcing amplitude $\delta=\delta(\eps)$ this reduction is possible. Following \citet{Ducimetiere_PRE_2025}, it is tempting to start by presuming that the size of the leading-order response can be determined from the linearised problem to be $\mathcal{O}(\delta/\eps)$. Although this estimate is correct in some cases, we shall see that in other cases, where a multi-modal response is excited, the magnitude of $\boldsymbol{q}$ can be different and determined only by appropriately balancing the nonlinear terms at various temporal frequencies beyond that of the forcing (e.g. see \S \ref{sec:PPF}).

\subsection{Towards a weakly nonlinear expansion} \label{sec:weak-nonlin-exp}

Returning to the full  forced nonlinear equation (\ref{eq:eq_main}), we now posit a weakly nonlinear, multiple time-scale expansion by introducing  a slow time $T :=\Delta\, t$,  where $\Delta \ll 1$ is a currently unspecified small parameter to be determined in terms of $\eps$, so that $\partial_t \mapsto \partial_t + \Delta\partial_T$. This `two-timing' anticipates a harmonic response on the `fast' forcing time scale $t$, with the fast-time-averaged amplitude of the solution evolving on the slow time scale $T$ due to weak nonlinearity. Substituting the ansatz
\begin{equation}
    \boldsymbol{q}(\boldsymbol{x},t,T) = \sum_{ n \in \mathbb{Z}}\boldsymbol{\hat{q}}_n(\boldsymbol{x},T) e^{\mathrm{i}n\omega t}
    \label{eq:slow-time-ansatz}
\end{equation}
(where $\bhq_{-n} = \bhq^*_{n}$, the complex conjugate of $\bhq_n$, to maintain a real solution) into equation (\ref{eq:eq_main}) and Fourier transforming in time gives
\beq
\Delta \partial_T \bhq_n + (\mathrm{i} n \omega I - \opL) \bhq_n 
= \big\{\N(\bhq,\bhq)\big\}_n + \delta{\boldsymbol{\hat{f}}}\,\delij_{n,\pm 1}
\label{eq:each-n}
\eeq
where 
\beq
    \big\{\N (\bhq,\bhq)\big\}_n = 
    \sum_{k\in\mathbb{Z}}  \N(\bhq_{n-k},\bhq_k).
\eeq
and $\delij_{ij}$ is the usual Kronecker delta function. 
Introducing some form of spatial discretization so $\bhq_n \in \mathbb{C}^M$ for integer $M$, 
the resolvent operator is most naturally represented by a singular value decomposition
\begin{equation}
\mathcal{R}(\mathrm{i} n \omega) = U_n \Sigma_n V_n^{{H}}
=
    \begin{pmatrix}
        \vert & & \vert\\
        \boldsymbol{{u}}_n^{(1)} & \ldots & \boldsymbol{{u}}_n^{(M)} \\
        \vert &  & \vert 
    \end{pmatrix}
    \begin{pmatrix}
        \sigma_n^{(1)} &  & \\
         & \ddots &    \\
         &  &  \sigma_n^{(M)} 
    \end{pmatrix}
    \begin{pmatrix}
        \vert & & \vert\\
        \boldsymbol{{v}}_n^{(1)} & \ldots & \boldsymbol{{v}}_n^{(M)} \\
        \vert &  & \vert 
    \end{pmatrix}^H,
\end{equation}
where $\Sigma$ is a diagonal matrix in $\mathbb C^{M \times M}$ whose entries are the real, positive singular values ordered such that  $\sigma_n^{(1)} \geq \sigma_n^{(2)} \geq \dots \geq \sigma_n^{(M)} \geq 0$, $U_n$ and $V_n$ are unitary $\mathbb C^{M \times M}$ matrices  whose $j$th columns are referred to as the $j$th left and right singular vectors respectively, and $(\,\cdot\,)^H$ denotes the Hermitian (conjugate) transpose.  Since $U_n$ and $V_n$ are unitary, their columns form an orthonormal basis with respect to the Hermitian inner product 
\beq
\langle \boldsymbol{z},\boldsymbol{w}\rangle := \sum_i z_i^* w_i
\eeq
for two complex vectors $\boldsymbol{z}, \boldsymbol{w} \in \mathbb{C}^M$. Significantly, for what follows, nothing {\it a priori} is known about how  these orthonormal bases project onto each other unless $\R$ is essentially rank-1 (see appendix \ref{Gelfand}). In particular, if the singular values have some non-trivial scaling with $\eps$, the projection of $\bv_n^{(i)}$ onto $\bu_n^{(j)}$ will typically also depend on $\eps$, which significantly complicates the weakly nonlinear analysis as we illustrate in what follows. 
The largest singular value, $\sigma_n^{(1)}$, represents the maximum possible linear response to forcing at frequency $\mathrm{i} n \omega$ so
\beq
\sigma_n^{(1)}:=\|\mathcal{R}(\mathrm{i} n \omega)\|
\eeq
with the corresponding leading right ($\bv_n^{(1)})$ and left ($\bu_n^{(1)}$) singular vectors representing the optimal forcing and response, respectively. The definition (\ref{highly}) then means
\beq
\eps:=\frac{1}{\sigma_1^{(1)}} \ll 1.
\eeq
(Later this may be relaxed to just $\eps=O(1/\sigma_1^{(1)})$ when a slightly more natural definition is available; e.g. see (\ref{highly2}) in \S \ref{sec:PPF}\,).
We proceed by expanding $\bhq_n$ in the orthonormal basis spanned by the left singular vectors associated with the linear response at each frequency $n$
\begin{equation}
    \bhq_n(T) =  \sum_{j=1}^M a_n^{(j)}(T)\boldsymbol{{u}}_n^{(j)}
\end{equation}
and then project equations (\ref{eq:each-n}) onto the hierarchy of forcing structures $\boldsymbol{v}_n^{(j)}$.  Doing so extracts each amplitude $a_n^{(j)}$ at the linear level due to the orthogonality of the singular vectors
\beq
\big\langle\boldsymbol{v}_n^{(j)},(\mathrm{i}n\omega I - \mathcal{L})\bhq_n \big\rangle 
=  \sum_{k=1}^M \frac{1}{\sigma_n^{(k)}}
\big\langle
\boldsymbol{v}_n^{(j)},\boldsymbol{{v}}_n^{(k)}
\big\rangle\big\langle
\boldsymbol{{u}}_n^{(k)},\bhq_n 
\big\rangle
= \frac{1}{\sigma_n^{(j)}}a_n^{(j)}
\eeq
(with no repeated summation of $n$ or $j$).
Applying this projection to the full nonlinear problem reduces (\ref{eq:each-n}) to a system of ordinary differential equations governing the slow-time evolution of each amplitude $a_n^{(j)}$ (so the equations are indexed by $n=0,1,\ldots,N$, where $N$ represents a maximum cut-off frequency of $N\omega$, and by $j=1,2,\ldots, M$):
\beq
\label{eq:final_gen_eq}
    \Delta \sum_{k}\Ov_n^{jk} \partial_T a_n^{(k)}    
    + \frac{1}{\sigma_n^{(j)}}a_n^{(j)} = 
    \sum_{m,k,\ell} \NL_{n-m,m}^{jk\ell}\,a_{n-m}^{(k)}\, a_m^{(\ell)} + \delta \langle\boldsymbol{v}_n^{(j)}, {\boldsymbol{\hat{f}}}\,\rangle \delij_{n,\pm 1},
\eeq
where the projection and nonlinear coefficients are, respectively,
\beq
        \Ov_n^{jk} := \langle\boldsymbol{v}_n^{(j)}, \boldsymbol{u}_n^{(k)}\rangle 
        \quad \& \quad    
        \NL_{n,m}^{jk\ell} := \big\langle\boldsymbol{v}_{n+m}^{(j)},\,\mathcal{N}\big(\boldsymbol{u}_n^{(k)},\boldsymbol{u}_m^{(\ell)}\big)\big\rangle.
\label{eq:coefficients}
\eeq
One of the main difficulties in what follows is discerning how both of these families of coefficients depend on $\eps$ in the limit $\eps \rightarrow 0$ so  that a rationally-consistent hierarchy of equations can be derived. In classical WNA, the equivalent of these coefficients would be {\em independent} of $\eps$, there defined as a distance from the bifurcation point, making this task straightforward in that setting. Furthermore, in classical WNA of, e.g., a pitchfork bifurcation, $\sigma_1^{(1)}$ would be infinite as the linear operator is singular for the bifurcating eigenfunction while all other singular values are $\mathcal{O}(1)$, i.e all the other linear operators are readily invertible. Here, however, the singular values generally all will have non-trivial scalings with $\eps$ since there is nothing particularly special about $\sigma_1^{(1)}$. This issue can manifest itself in two different forms: (i) multiple singular values at the same frequency scaling similarly, e.g. for the forcing frequency $\sigma_1^{(j)}=\mathcal{O}(\sigma_1^{(1)})$ for a range of integer $j \geq 2$; and (ii) leading singular values across different frequencies scaling similarly, i.e. $\sigma_n^{(1)}=\mathcal{O}(\sigma_1^{(1)})$, or even dominating that at the forcing frequency, i.e. $\sigma_1^{(1)}=o(\sigma_n^{(1)})$ for some integer $n \neq 1$. The last situation looks particularly pathological but in fact arises in channel flow, along with scenario~(i) (both for $n=0$) and is our first example below in \S \ref{sec:PPF}.

Despite these various new complications, the hope is still to obtain a  large dimensionality reduction to a small system of amplitude equations by leveraging the strong non-normality in the problem. Given the discussion above, this reduction has to be approached by carefully determining how each of the terms in equations (\ref{eq:final_gen_eq}) depends on $\eps$, which is generally system dependent.  The amplitude equation discussed by \citet{Ducimetiere_PRE_2025} imposes a pre-defined asymptotic structure for the solution amplitudes  and the scaling of these coefficients mimicking the standard WNA approach (see \S\ref{sec:duc_compare} for further detail). We shall see in the following sections that the asymptotic reduction needed here  for a forced highly non-normal operator is far richer than this, with the intrinsic weakly nonlinear response of these systems sometimes even being multi-modal at leading order.

%
%
\section{Plane Poiseuille flow } \label{sec:PPF}

We begin by considering  plane Poiseuille flow (PPF), which was initially analysed by \citet{ducimetiere2022weak}. 
In this flow, a Newtonian fluid confined between two fixed fixed plates at $y = \pm h$ is driven by an imposed pressure gradient. The flow in this channel is assumed to be periodic in the streamwise $x$ and spanwise $z$ directions.  As is customary, the Navier--Stokes equations are non-dimensionalised using the channel half-height $h$, the laminar centreline velocity $U_{max}$ and pressure $\rho U_{max}^2$.  In the absence of harmonic forcing, the laminar flow driven by a dimensionless pressure gradient $\partial_xP = -2/\Rey$ is then given by  $\boldsymbol{U} = (1-y^2)\boldsymbol{e}_x$, where $\boldsymbol{e}_x$ is the streamwise unit vector, $\Rey := U_{max}h/\nu$ and $\nu$ is the kinematic viscosity.  Fluctuations about the laminar state are described by the following harmonically-forced equations:
\begin{subequations}
\label{eq:Navier_stokes_PPF}
\begin{alignat}{1}
    \partial_t\boldsymbol{u} = -\boldsymbol{U}\bcdot\bnabla\boldsymbol{u} - \boldsymbol{u}\bcdot\bnabla\boldsymbol{U} - \bnabla p + &\frac{1}{\Rey}\nabla^2\boldsymbol{u} - \boldsymbol{u}\bcdot\bnabla\boldsymbol{u} + \delta({\boldsymbol{\hat{f}}}e^{\mathrm{i}(k_x x +\omega t)} + {c.c.}), \\
    &\bnabla\bcdot\boldsymbol{u} = 0.
\end{alignat}
\end{subequations}
To compare directly to the results in \S2.2 of \citet{ducimetiere2022weak}, we fix the wavenumbers of the forcing to $(k_x,k_z) = (-1.2,0)$, which ensures the linear stability of the laminar profile for all $\Rey$ \citep{lin1946stability,Lin1955book,deguchi2018bifurcation}. (Note that the phase velocity $c:=-\omega/k_x$ needs to be $\in (0,1)$ to obtain a large response, and we take $\omega>0$ in the linear problem.) The non-normality in this 2-dimensional configuration is associated strictly with the Orr mechanism \citep{Orr-07,Farrell_orr}, where perturbations initially inclined upstream are amplified as they are tilted downstream by the mean shear.

%
%
\subsection{Linear problem: resolvent analysis} \label{sec:linear_PPF}

Equations (\ref{eq:Navier_stokes_PPF}), linearised about the laminar flow, have been solved numerically using the open-source pseudo-spectral framework Dedalus \citep{dedalus_methods}.  In particular, resolvent analysis can be performed efficiently following the method of \citet{skene2026fast}, where matrix-free methods using automated adjoints are easily utilised to perform a sparse singular value decomposition of the resolvent.  The nonlinear interaction of the resulting singular vector fields can then be evaluated using built-in differential operators in Dedalus.  This allows the analysis to be performed quickly and at high resolution on a laptop computer, with the implementation being flexible to changes in the governing equations or the geometry of the domain.  The method has been validated using an in-house code, and the results are commensurate with numerical values reported by \citet{ducimetiere2022weak}.  Critical layers develop in the singular vectors of the resolvent at high $\Rey$, as in the eigenfunctions of the Orr--Sommerfeld equations \citep{MasloweSA1986CLiS}, and thus a large number of modes is needed in the wall-normal direction.  In the results presented, the wall-normal direction is discretised using Chebyshev polynomials, typically employing $N_y = 512$ modes in the vertical direction, but tests with up to $N_y=4096$ have also been conducted with no significant changes in the scaling with $\varepsilon$ of the various terms.\\

%
%
\begin{figure}
    \captionsetup{width=\columnwidth}
    \centering
    \includegraphics[width=1\linewidth]{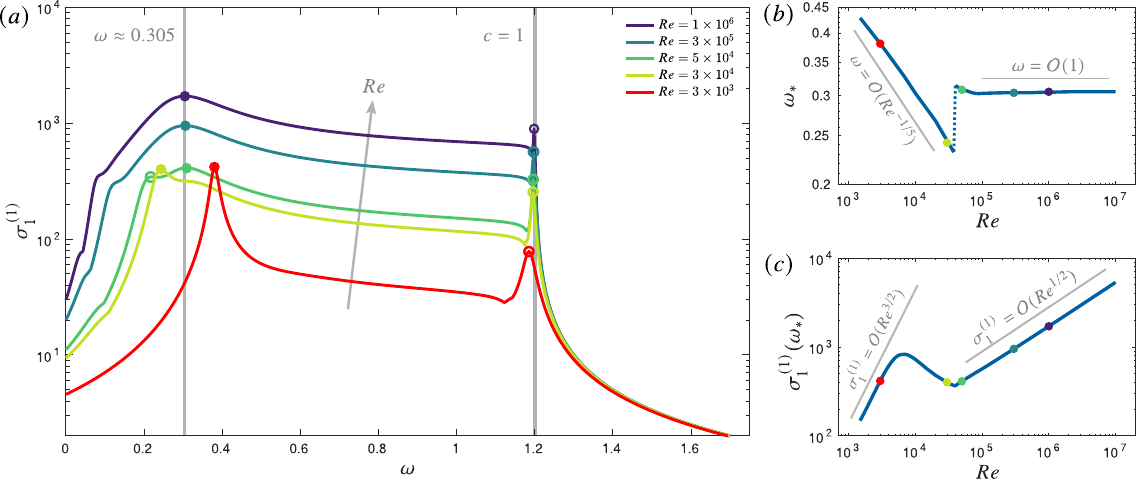}
    \caption{$(a)$ Leading singular value of the resolvent $\mathcal{R}(\mathrm{i}\omega)$ with $(k_x,k_z)=(-1.2,0)$ for varied forcing frequency $\omega$ at values of $\Rey$ given in the legend.  Local maxima are marked with open circles, with the global maximum at each $\Rey$ marked with filled circles.  $(b)$ Optimal forcing frequency $\omega_*$ at each $\Rey$ and $(c)$ the corresponding leading singular value $\sigma_1^{(1)}$.  Markers in $(b,c)$ correspond to the global maxima in $(a)$.}
    \label{fig:sigma_scaling_PPF}
\end{figure}
To identify a regime appropriate for the asymptotic reduction, we begin by examining how the leading singular value, $\sigma_1^{(1)}$, of the resolvent $\mathcal{R}(\mathrm{i} \omega)$ varies as a function of the forcing frequency $\omega$; see figure \ref{fig:sigma_scaling_PPF}$(a)$.  At low $\Rey$ (e.g. $\Rey=3 \times 10^3$ where \citet{ducimetiere2022weak} operated), $\sigma_1^{(1)}$ has two local maxima as a function of $\omega$; however, as $\Rey$ is increased, a third local maximum begins to appear (see the $\Rey=5 \times 10^4$ curve), which develops into the asymptotic global maximum of the system.  This behaviour is made clear by examining the optimal forcing frequency $\omega_*$ for a given $\Rey$
\begin{equation}
    \omega_* =\argmax_\omega \big(\sigma_1^{(1)}(\omega;\Rey)\big),
\end{equation}
which is plotted as a function of $\Rey$ in figure \ref{fig:sigma_scaling_PPF}$(b)$.  The corresponding singular values are shown in figure \ref{fig:sigma_scaling_PPF}$(c)$.  
There is a discrete jump in $\omega_*$ as the global maximum switches between these local maxima, which corresponds to a change in scaling of $\sigma_1^{(1)}$ with $\Rey$. 
This distinguishes two different regimes of behaviour: (i) a pre-asymptotic regime where $\omega = \mathcal{O}(\Rey^{-1/5})$ when $\Rey\lesssim 4\times10^4$ and $\sigma_1^{(1)} = \mathcal{O}(\Rey^{3/2})$ for at least $\Rey \lesssim 7 \times 10^3$, and (ii) the asymptotic regime where $\sigma_1^{(1)} = \mathcal{O}(\Rey^{1/2})$ with $\omega = \mathcal{O}(1) \approx 0.305$ when $\Rey\gtrsim 4\times10^4$.  An asymptotically-consistent reduction of this system can only be attempted for  regime (ii) as only there can $\eps$ be made as small as desired by taking $\Rey\rightarrow\infty$ (ignoring the off-topic `resonance' limit in which $k_x$ and $\omega$  are manipulated carefully to approach the imaginary eigenvalue on the neutral curve of the linear operator).
In contrast, $\sigma_1^{(1)}$ actually begins to {\it decrease} as $\Rey$ is increased ($\Rey \gtrsim 7\times10^3$) in regime (i) despite the optimal frequency $\omega_*$ varying smoothly. No rational asymptotic expansion can be attempted in this finite range of $\Rey$ as $\eps$ always stays bounded away from zero.

\begin{figure}
    \captionsetup{width=\columnwidth}
    \centering
    \includegraphics[width=1\linewidth]{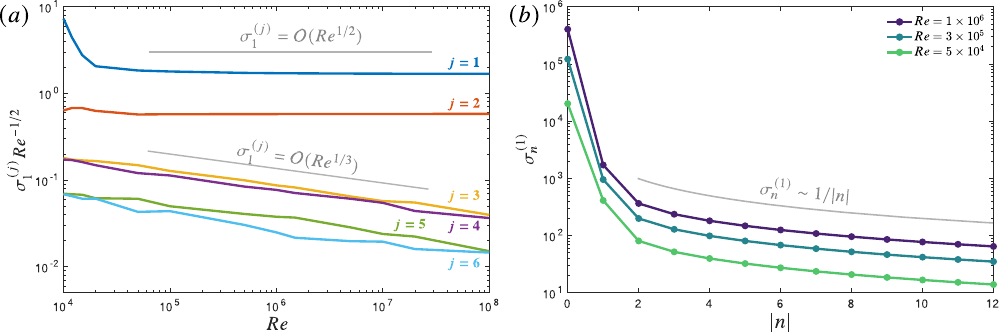}
    \caption{$(a)$ Compensated scaling of the first six singular values $\sigma_1^{(j)}$ $1 \leq j \leq 6$ at the forcing frequency $\omega=0.305$. $(b)$ Leading singular value for each frequency $n\omega$ ($\omega=0.305$) with streamwise wavenumber $nk_x$ $(k_x=-1.2)$.}
    \label{fig:sigma_subdom_scaling_PPF}
\end{figure}

Focusing on regime (ii), we formally define $\eps$ as
\begin{equation}
    \eps := \Rey^{-1/2}=\mathcal{O}\biggl( \frac{1}{\sigma_1^{(1)}}\biggr),
    \label{highly2}
\end{equation}
in the limit as $\Rey\rightarrow\infty$ with the forcing frequency $\omega=\mathcal{O}(1) \approx0.305$.  This is slightly different from the equality used in (\ref{highly}) but is more convenient here given the presence of the underlying Reynolds number controlling $\eps$.
The task now is to determine the scaling of the various terms in equation (\ref{eq:final_gen_eq}) with $\varepsilon$.  
We begin by analysing the various singular values $\sigma_n^{(j)}$, numerically obtaining
\begin{equation}
\sigma_n^{(j)} =
    \begin{cases}
        \: \mathcal{O}(\Rey^{1/2}), & j \in \{1,2\}, \\
        \: \mathcal{O}(\Rey^{1/3}), & j\geq 3,
    \end{cases}
    \label{eq:leading_sin_val_PPF}
\end{equation}
for $|n|\geq 1$, as shown in figure~\ref{fig:sigma_subdom_scaling_PPF}$(a)$ for $n=1$. (Note that $\sigma_n^{(j)}$ is the $j^{th}$ singular value for frequency $n \omega$ and correspondingly the streamwise wavenumber is taken as $n k_x$; see figure~\ref{fig:sigma_scaling_PPF}.)  Significantly, the two leading singular values $(j=\{1,2\})$ are asymptotically separated from the remaining subdominant singular values.  The magnitudes of the singular values typically decrease with $|n|$ as shown for $j=1$ in figure \ref{fig:sigma_subdom_scaling_PPF}$(b)$ at various $\Rey$.  In each case $\sigma_n^{(1)} \sim 1/|n|$ for $|n|\gtrsim2$.
%

%
%

%
%
\begin{bottomfigure}
    \captionsetup{width=\columnwidth}
    \centering
    \includegraphics[width=1\linewidth]{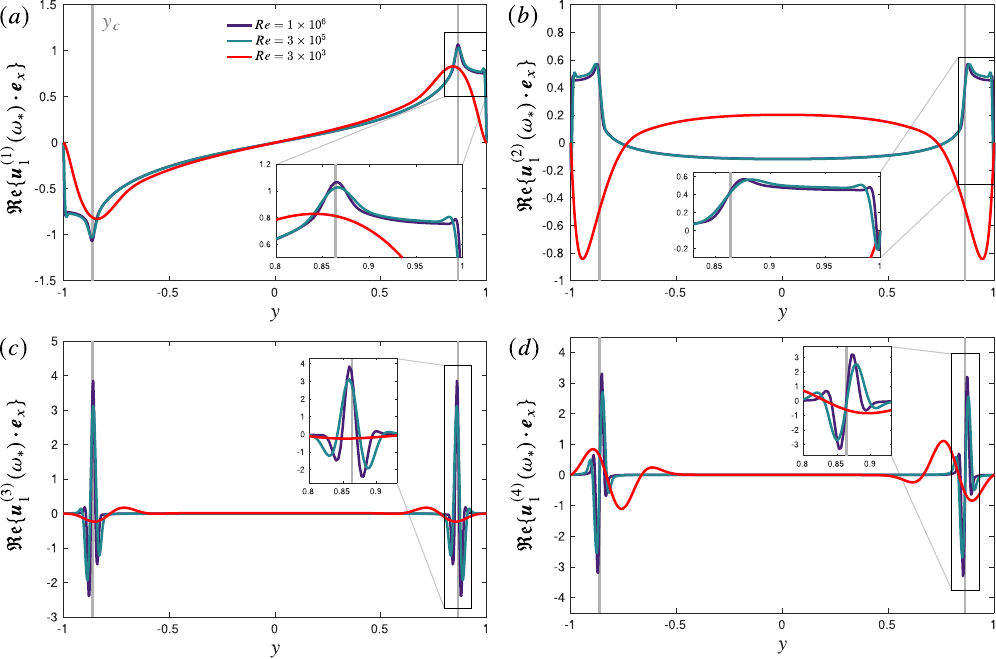}
    \caption{Real part of the streamwise component of the $(a)$ leading $\bu_1^{(1)}$, $(b)$ secondary $\bu_1^{(2)}$, $(c)$ tertiary $\bu_1^{(3)}$, and $(d)$ quaternary $\bu_1^{(4)}$ left singular vectors as a function of the wall-normal direction $y$.  The insets show a zoomed-in view.  The location of the critical layers $y_c$ are shown by the vertical grey lines.}
    \label{fig:u_svd_modes_PPF}
\end{bottomfigure}

%
%
\begin{figure}
    \captionsetup{width=\columnwidth}
    \centering
    \includegraphics[width=1\linewidth]{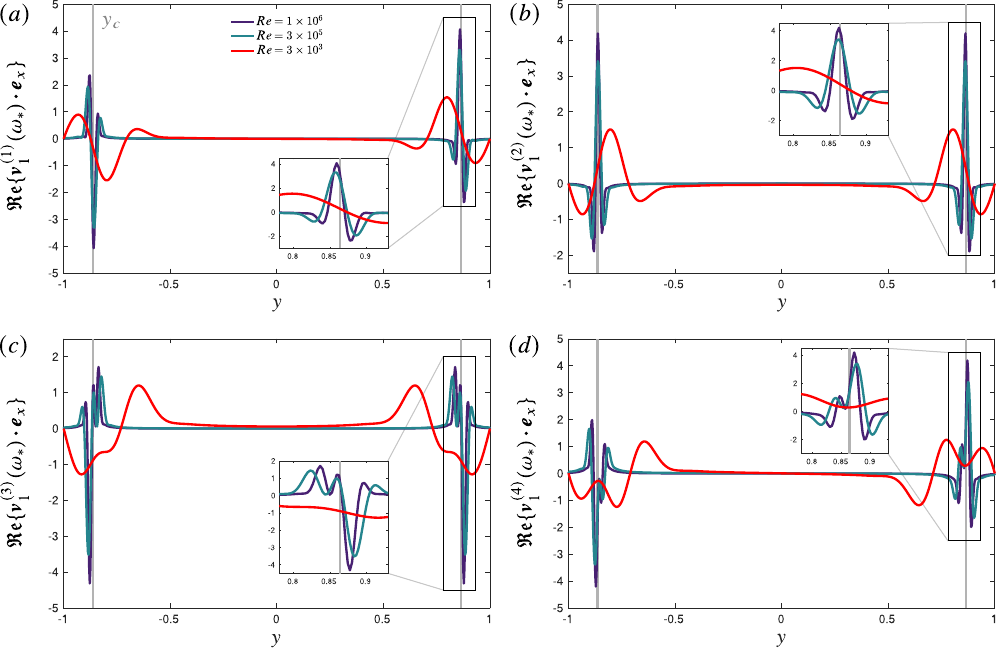}
    \caption{Real part of the streamwise component of the $(a)$ leading $\bv_1^{(1)}$, $(b)$ secondary $\bv_1^{(2)}$, $(c)$ tertiary $\bv_1^{(3)}$, and $(d)$ quaternary $\bv_1^{(4)}$ right singular vectors as a function of the wall-normal direction $y$.  The insets show a zoomed-in view.  The location of the critical layers $y_c$ are shown by the vertical grey lines.}
    \label{fig:v_svd_modes_PPF}
\end{figure}

The real parts of the streamwise component of the leading left singular vectors for $n=1$ are shown  in figure \ref{fig:u_svd_modes_PPF}$(a,b)$ for $j = \{1,2\}$ together with equivalent plots in $(c,d)$ for $j=\{3,4\}$.
The corresponding singular vector fields are all 2-dimensional (recall $k_z=0$) and are arranged into pairs having a symmetric/antisymmetric  streamwise component ($j\,$=\,2/$j$\,=\,1 and $j$\,=\,3/$j$\,=\,4)  about the midplane.  At high $\Rey$, the singular modes develop critical layers at $y_c=\pm\sqrt{1+n\omega/k_x}$ as shown by the grey lines for $n$\,=\,1 in figure \ref{fig:u_svd_modes_PPF}, with the critical layer behaviour dominating the interior for $j=3,4$ (and higher $j$ not shown) whereas the opposite is true for $j=1,2$ (an investigation of  this feature is underway but is beyond the scope of this current report). Figure \ref{fig:v_svd_modes_PPF}$(a,b)$ shows the corresponding right singular vectors, which are all dominated by structures at the critical layers.  
Determining the behaviour of these singular vector fields in the critical layer will be crucial for rationalising the observed scaling of the various coefficients that appear in the amplitude equations \eqref{eq:final_gen_eq}.  If strictly localised in the critical layer (e.g. any $\bv_{n\neq0}^{(j)}$ or $\bu_{n\neq0}^{(j>2)}$), the singular vectors exhibit a $\mathcal{O}(\Rey^{1/6})$ scaling as a consequence of the unit normalisation of these vectors
and the critical layer having a thickness $\mathcal{O}(\Rey^{-1/3})$ (as is necessary for the viscous terms to enter the balanced equations; see \citet{MasloweSA1986CLiS}).  However, when activity is observed outside the critical layer (e.g. $\bu_{n\neq0}^{(j\leq2)}$), the singular vectors remain $\mathcal{O}(1)$ across $y$.

The numerical singular value scalings listed in (\ref{eq:leading_sin_val_PPF}) exclude the special case $n=0$ for which $(k_x,k_z)=(0,0)$.
Here the linear operator for the velocity simply reduces to the Laplacian, i.e. $\opL = \Rey^{-1}\nabla^2$, acting on $u:=\bu\bcdot\boldsymbol{e}_x$ (since in 2D incompressibility and boundary conditions force $v:=\bu\bcdot\boldsymbol{e}_y=0$ everywhere), which, with no-slip boundary conditions, is self-adjoint. It then follows that the the left and right singular vectors coincide with the eigenvectors of $\opL$, and the singular values = $-1/\lambda_j$, where the eigenvalues $\lambda_j$ of $\opL$ are all real and negative.  The relevant singular vectors are then of the form $\bu_0^{(j)}=\bv_0^{(j)}=(u_0^{(j)},0)$, with
\beq
\sigma_0^{(j)} = \frac{4\Rey}{j^2\pi^2}, \qquad
u_0^{(j)} = \Biggl\{  \begin{array}{ll}
\cos\Big(y \sqrt{\Rey/\sigma_0^{(j)}}\Big), & j = 1,3,5,\dots  \\
\sin\Big(y \sqrt{\Rey/\sigma_0^{(j)}} \Big), & j= 2,4,6,\dots
\end{array}
\biggr.
\label{eq:mean_singular_values}
\eeq
so that odd/even $j$ are symmetric/antisymmetric about the midplane $y=0$.

Given the definition of $\eps$ in (\ref{highly2}), 2D channel flow therefore has the singular value properties
\begin{align}
\sigma_{n \neq 0}^{(j)} &= \biggl\{\begin{array}{ll}
        \: \mathcal{O}(\varepsilon^{-1}),   & j \in \{1,2\}\\
        \: \mathcal{O}(\varepsilon^{-2/3}), & j\geq 3
                             \end{array} \biggr.
\nonumber \\
\sigma_0^{(j)} &= \,\,\,\,\,\,\,\, \mathcal{O}(\varepsilon^{-2}), \qquad \forall j.
\label{eq:PPF_singular_val_scaling}
\end{align}
This scaling exhibits both the issues discussed below equation (\ref{eq:coefficients}): $\sigma_1^{(1)}$ is subdominant to $\sigma_0^{(1)}$ {\it and} $\sigma_0^{(j)} = \mathcal{O}( \sigma_0^{(1)} )$ for all $j\geq 2$ as $\eps \rightarrow 0$. 
To give a flavour of the implications, note that in classical WNA if the leading mode $\bu_1$ has an amplitude $\eps$, the nonlinear self-interaction term  $\N(\bu_1, \bu_1)$ would be $\mathcal{O}(\eps^2)$, which forces a mean and second harmonic at the same higher-order magnitude $\mathcal{O}(\eps^2)$. In  `non-normal' WNA, however, an $O(\eps^2)$ nonlinear term could generate an $\mathcal{O}(1)$ mean response due to the large singular value $\sigma_0^{(1)}=\mathcal{O}(1/\eps^2)$ magnifying the mean response. This nonlinearly forced mean-flow component then dominates the presumed leading (harmonically-forced) term, implying an inconsistency. 

%
%
\subsection{Weakly nonlinear reduction in the asymptotic regime} \label{sec:nonlinear_red_PPF}
A key observation for deriving an asymptotically reduced system in this case is that the amplitudes of the first two ($j\in \{1,2\}$) left singular vectors for any $n \neq 0$ should dominate the rest ($j \geq 3$) because of their asymptotically larger singular values (see (\ref{eq:PPF_singular_val_scaling})). This implicitly assumes  the absence of any anonymously larger (i.e. $\mathcal{O}(\eps^{1/3})$ larger) nonlinear forcing of a $j\geq 3$ singular vector compared to either $j$\,=\,1 or $j$\,=\,2, which holds true here and probably more generally (\,i.e. $\NL_{n,m}^{j\geq 3 \,k\ell}$ is $\mathcal{O}(\NL_{n,m}^{1k\ell})$ or $\mathcal{O}(\NL_{n,m}^{2k\ell})$\,). With this ordering, the analysis greatly simplifies as only these two amplitudes need to be considered for each $n \neq 0$. 

To determine the reduced equations, it is necessary to evaluate the scaling of the various coefficients that appear in equation \eqref{eq:final_gen_eq} with $\Rey$, which have been examined numerically.  For $n\neq0$, the projection integrals $\Ov_n^{jk}$ behave as
\beq
\Ov_n^{11}, \Ov_n^{22} = \mathcal{O}( \eps^{1/3} ) \quad \& \quad \Ov_n^{12}=\Ov_n^{21}=0,
\eeq
the latter of which is explained by the $y$-symmetry of the singular vector fields.  The scaling of the $\Ov_n^{jj}$ coefficients, however, is only explained by examining the $y$-structure of the singular vectors more closely.  Since $\bu_{n\neq0}^{(j\leq2)}$ remains $\mathcal{O}(1)$ across $y$, multiplication with $\bv_{n\neq0}^{(j)}$ maintains a critical-layer localised field that scales as $\mathcal{O}(\Rey^{1/6})$.  Thus, integrating this field across the $\mathcal{O}(\Rey^{-1/3})$ critical layer gives $\Ov_n^{jj} = \mathcal{O}(\Rey^{-1/6}) = \mathcal{O}(\varepsilon^{1/3})$.

%
%
\begin{figure}
    \captionsetup{width=\columnwidth}
    \centering
    \includegraphics[width=1\linewidth]{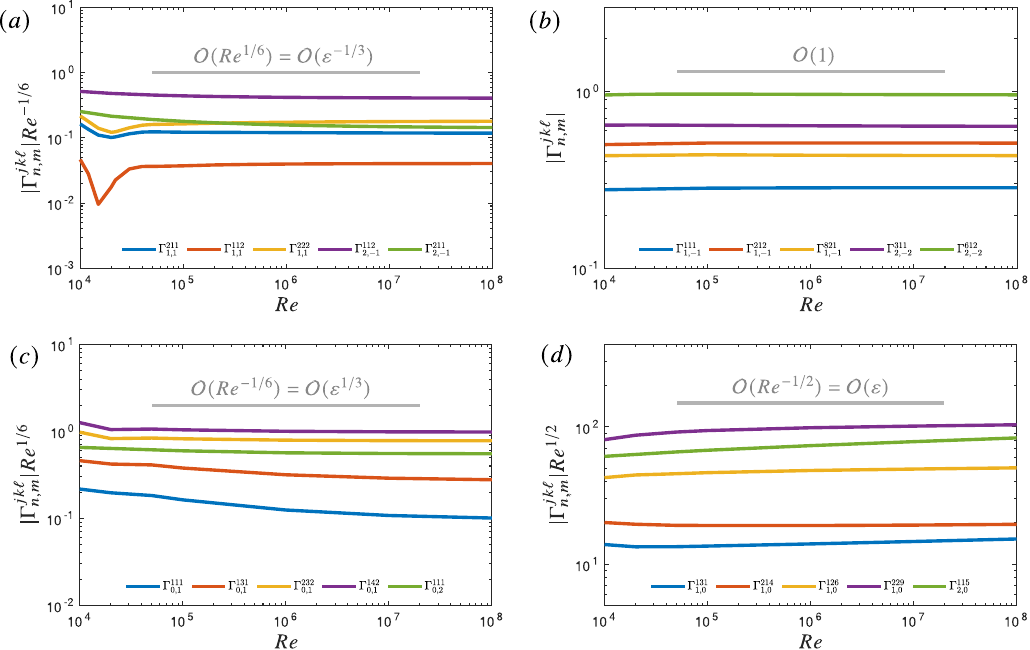}
    \caption{Compensated scalings of various nonlinear terms $\NL_{n,m}^{jk\ell}$ with $\Rey$.}
    \label{fig:nonlin_scaling_PPF}
\end{figure}

A similar approach can be taken to understand the scaling of the nonlinear coefficients $\NL_{n,m}^{jk\ell}$, which can be categorised into four different sets by their scaling with $\Rey$, as shown by the four subplots of figure \ref{fig:nonlin_scaling_PPF}.  The nonlinear interaction of the leading $n\neq0$ modes is dominated by the streamwise component $\N(\bu_{n\neq0}^{(k\leq2)}, \bu_{m\neq0}^{(\ell\leq2)})\bcdot \boldsymbol{e}_x = \mathcal{O}(\partial_y\bu_{n\neq0}^{(\ell\leq2)}\bcdot\boldsymbol{e}_x) = \mathcal{O}(\Rey^{1/3})$ in the critical layer.  A subsequent integration against the right singular vectors distinguish two cases: (a) for $\bv_{n+m\neq0}^{(j)} = \mathcal{O}(\Rey^{1/6})$, integration across the $\mathcal{O}(\Rey^{-1/3})$ critical layer sets the nonlinear coefficient at $\mathcal{O}(\Rey^{1/6})=\mathcal{O}(\varepsilon^{-1/3})$ (figure \ref{fig:nonlin_scaling_PPF}$(a)$); and (b) for $\bv_0^{(j)} = \mathcal{O}(1)$, integration across the critical layer sets the nonlinear coefficient at $\mathcal{O}(1)$ (figure \ref{fig:nonlin_scaling_PPF}$(b)$).
When considering a nonlinear interaction between mean ($n=0$) and oscillatory ($n\neq0$ with $k$ or $\ell\le 2$) modes, we find that $\N(\bu_{n\neq0}^{(k)}, \bu_{0}^{(\ell)}) = \mathcal{O}(\N(\bu_{0}^{(k)}, \bu_{n\neq0}^{(\ell)})) = \mathcal{O}(1)$, and thus, anticipate that integration against $\bv_{n\neq0}^{(j)}$ would set the nonlinear coefficient at $\mathcal{O}(\Rey^{-1/6}) = \mathcal{O}(\varepsilon^{1/3})$, as is observed for $\NL_{0,n}^{jk\ell}$ in figure \ref{fig:nonlin_scaling_PPF}$(c)$.  However, a poor overlap in the final projection for the  $\NL_{n,0}^{jk\ell}$ coefficients result in a weaker scaling of $\mathcal{O}(\Rey^{-1/2}) = \mathcal{O}(\varepsilon)$ (figure \ref{fig:nonlin_scaling_PPF}$(d)$).

Using the determined scaling of the coefficients and accounting for the $y$-symmetries of the vector fields, gives the following form for the $n \neq 0$ equations for $a_n^{(1)}$ and $a_n^{(2)}$:
\begin{subequations}
\label{eq:oscillatory_mode_scaling}
\begin{align}
    \begin{split}
        (\ldots)\Delta \eps^{1/3} \partial_T a_n^{(1)} & \sim  (\ldots)  \eps a_n^{(1)} + \eps^{-1/3}\Big(\sum_{m\neq\{0,n\}}  (\ldots) a_{n-m}^{(1)}a_{m}^{(2)} \Big) \\
        &+ \eps^{1/3} \Big(\sum_{\ell\, {\rm odd}} (\ldots) a_0^{(\ell)} a_n^{(1)} +\sum_{\ell \, {\rm even}} (\ldots) a_0^{(\ell)}a_n^{(2)}\Big) + (\ldots)\delta \delij_{n,\pm 1},
    \end{split} \\
    \begin{split}
    (\ldots)\Delta \eps^{1/3} \partial_T a_n^{(2)} & \sim (\ldots)  \eps a_n^{(2)}+\eps^{-1/3}\Big(\sum_{m\neq\{0,n\}} (\ldots) a_{n-m}^{(1)}a_{m}^{(1)} + (\ldots) a_{n-m}^{(2)}a_{m}^{(2)}\Big) \\
    &+ \eps^{1/3} \Big(\sum_{\ell\, {\rm even}} (\ldots) a_0^{(\ell)} a_n^{(1)} + \sum_{\ell\, {\rm odd}} (\ldots) a_0^{(\ell)} a_n^{(2)}\Big)  + (\ldots) \delta \delij_{n,\pm 1},
    \end{split}
\end{align}
\end{subequations}
where $(\ldots)$ indicates an $\mathcal{O}(1)$ coefficient suppressed for clarity (e.g. the forcing is presumed not to be nearly orthogonal to $\boldsymbol{v}_1^{(j)}$).  

The situation for the mean ($n=0$) left singular vector amplitudes is not so clear cut  as their singular values have the same order for all $j$, which asymptotically dominate those for $n \neq 0$. However, the $n=0$ singular vectors only have a streamwise component and are strictly functions of $y$ so there is no nonlinear self-interaction of these modes.  Considering the symmetries of the singular vector fields and noting that 
\beq
\Ov_0^{jk} = \delij_{jk}
\eeq
as the left and right singular vectors are the same orthonormal set, the structure of the $n=0$ equations is
\begin{subequations}
\label{eq:mean_mode_scaling}
    \begin{alignat}{2}
    \Delta \partial_T a_0^{(\ell)} &\sim (\ldots) \eps^2 a_0^{(\ell)} + \,\,\,\,\,\,\sum_{m\neq0} (\ldots)a_m^{(1)}a_{-m}^{(2)}, \quad&& \ell \,\, {\rm odd}, \label{eq:mean_mode_scaling_odd}\\
    \Delta \partial_T a_0^{(\ell)} &\sim (\ldots) \eps^2 a_0^{(\ell)} + \sum_{m\neq0} 
    \Big( (\ldots) a_m^{(1)}a_{-m}^{(1)}  + (\ldots) a_m^{(2)}a_{-m}^{(2)}\Big), \quad&&\ell \,\, {\rm even}. \label{eq:even_mean}
    \end{alignat}
\end{subequations}
Balancing linear and forcing terms in (\ref{eq:oscillatory_mode_scaling}) means that $a_1^{(1)}$ and $a_1^{(2)}$ are $\mathcal{O}(\delta/\varepsilon)$.  Since we anticipate involvement of the mean amplitudes, the mean-fluctuation nonlinear interaction is balanced with the linear amplification in (\ref{eq:oscillatory_mode_scaling}) to obtain $a_0^{(\ell)} = \mathcal{O}(\varepsilon^{2/3})$.  The asymptotic size of $a_1^{(1)}$ and $a_1^{(2)}$ then can be obtained from the mean equations (\ref{eq:mean_mode_scaling}), which implies $\delta = \mathcal{O}(\varepsilon^{7/3})$ and $\Delta=\mathcal{O}(\eps^2)$. The latter scaling indicates that the slow time-derivative terms in (\ref{eq:oscillatory_mode_scaling}$a,b$) can be neglected.  Asymptotic consistency is checked by considering the size of the remaining nonlinear terms in (\ref{eq:oscillatory_mode_scaling}): fluctuation-fluctuation nonlinear interactions arise at the same order as the mean-fluctuation terms, yielding asymptotically balanced equations as $\varepsilon\rightarrow0$.
The result of all this is as follows.
\begin{my_indent}
    \item Mean modes have an asymptotic size $a_0^{(\ell)} = \mathcal{O}(\varepsilon^{2/3})$.
    \item Leading-order oscillating modes ($|n|\geq 1$) have an asymptotic size $a_n^{(j)} = \mathcal{O}(\varepsilon^{4/3})$ for $j=\{1,2\}$.  For $j \geq 3$, the amplitudes are subdominant and do not appear in the leading order equations.
    \item The forcing amplitude depends on $\varepsilon$ as $\delta = \mathcal{O}(\varepsilon^{7/3})$.
    \item The slow time-scale $\Delta = \mathcal{O}(\varepsilon^{2})$.
\end{my_indent}
In particular, the asymptotic solution takes the form
\begin{equation}\begin{split}
    \boldsymbol{u}(x,y,t) &=  \eps^{2/3} 
    \sum_{j}  A_0^{(j)}(\varepsilon^2 t) \,\boldsymbol{u}_0^{(j)}(y) \nonumber \\
    &+\eps^{4/3} \Big(\sum_{n \neq 0} \Bigl[A_n^{(1)}(\varepsilon^2 t) \bu_n^{(1)}(y)+A_n^{(2)}(\varepsilon^2 t) \bu_n^{(2)}(y) \Bigr]\,e^{\mathrm{i}n(k_xx+\omega t)} + c.c.\Big) + \mathcal{O}(\varepsilon^{5/3})
    \label{eq:PPF_solution}
    \end{split}
\end{equation}
where 
\begin{equation}
    a_0^{(j)} = \eps^{2/3}A_0^{(j)}, \quad {\&} \quad a_n^{(j)} = \eps^{4/3}A_n^{(j)} \quad j \in\{1,2\}
\end{equation}
and $A_{-n}^{(j)}=A_n^{(j)*}$ to ensure realness of the solution.  This solution is dominated by the mean response and the non-normal WNA is only possible here in 2D channel flow if the forcing $\delta= \mathcal{O}(\eps^{7/3})$ as $\eps \rightarrow 0$.

To write down a final set of $\eps$-independent amplitude equations, the scaling with $\eps$ is extracted from each coefficient to render it $\mathcal{O}(\eps^0)$ and a $\hat{\cdot }$ is added to the coefficient to  indicate this, e.g.
\beq
\NL^{112}_{n-m,m}= \eps^{-1/3} \hat{\NL}^{112}_{n-m,m} \qquad m \not \in\{0,n\} \,\,\& \,\, n \neq 0.
\eeq
In a similar vein,
\beq
    \sigma_0^{(j)} = \frac{ \hat{\sigma}_0^{(j)} }{\eps^2}, 
    \quad {\rm and} \quad 
    \sigma_n^{(j)} =  \frac{ \hat{\sigma}_n^{(j)} }{\eps} \quad j \in \{1,2\} \,\,\& \,\, n \neq 0
\eeq
with $\delta = \varepsilon^{7/3}\hat{\delta}$.  Thus, the rescaled WNA equations to be solved are 
\begin{subequations}
\label{eq:PPF_reduced_sys}
\begin{align}
    &\qquad\qquad\partial_T{A}_0^{(\ell)} + \frac{1}{\hat{\sigma}_0^{(\ell)}}A_0^{(\ell)} = \sum_{m\neq0} \big(\hat{\NL}_{m,-m}^{\ell 12}+ \hat{\NL}_{m,-m}^{\ell 21}\big)\, A_{m}^{(1)}{A}_{-m}^{(2)}, \qquad \ell \,\, {\rm even} 
    \\
    &\begin{aligned}
        \partial_T{A}_0^{(\ell)} + \frac{1}{\hat{\sigma}_0^{(\ell)}}A_0^{(\ell)} = \sum_{m\neq0}\Big( \big(\hat{\NL}_{m,-m}^{\ell 11}+ &\hat{\NL}_{m,-m}^{\ell 11}\big)\, A_{m}^{(1)}{A}_{-m}^{(1)} \\
        &+ \big(\hat{\NL}_{m,-m}^{\ell 12}\hat{\NL}_{m,-m}^{\ell 22}\big)\, A_{m}^{(2)}{A}_{-m}^{(2)}\Big), \qquad \ell \,\, {\rm odd} 
    \end{aligned} \\
    &\begin{aligned}
        \frac{1}{\hat{\sigma}_n^{(1)}}A_n^{(1)} = \!\!\sum_{m\neq\{0,n\}}\!\!\Big(&\hat{\NL}_{n-m,m}^{112}\, A_{n-m}^{(1)}{A}_{m}^{(2)} + \hat{\NL}_{n-m,m}^{121}\, A_{n-m}^{(2)}{A}_{m}^{(1)}\Big)  \\
        &+ \sum_{\ell \, {\rm odd}} \hat{\NL}_{0,n}^{1\ell1}A_0^{(\ell)}A_{n}^{(1)}
        + \sum_{\ell \, {\rm even}} \hat{\NL}_{0,n}^{1\ell2}A_0^{(\ell)}A_{n}^{(2)} 
        + \langle \bv_n^{(1)}, \boldsymbol{\hat{f}}\rangle \hat{\delta} \delij_{n,\pm 1},
    \end{aligned} \\
    &\begin{aligned}
        \frac{1}{\hat{\sigma}_n^{(2)}}A_n^{(2)} = \!\!\sum_{m\neq\{0,n\}}\!\!\Big(&\hat{\NL}_{n-m,m}^{211}\, A_{n-m}^{(1)}{A}_{m}^{(1)} + \hat{\NL}_{n-m,m}^{222}\, A_{n-m}^{(2)}{A}_{m}^{(2)}\Big) \\
        &+ \sum_{\ell\, {\rm even}}\hat{\NL}_{0,n}^{2\ell1}A_0^{(\ell)}A_n^{(1)} 
        + \sum_{\ell\, {\rm odd}}\hat{\NL}_{0,n}^{2\ell2}A_0^{(\ell)}A_n^{(2)}
        + \langle \bv_n^{(2)}, \boldsymbol{\hat{f}}\rangle \hat{\delta} \delij_{n,\pm 1},
    \end{aligned}
\end{align}
\end{subequations}
for $n\neq0$.  This asymptotically-reduced system is still infinite-dimensional in time ($n \in \mathbb{Z}\setminus \{0\}$) and, for the mean ($n=0$) modes, formally all $\ell \leq M$ must be included, where $M$ is the large spatial dimension used to initially discretize the problem (in $y$). Fortunately, however, the singular values $\sigma_0^{(\ell)}$ decrease quickly for increasing $\ell$ and also (albeit less quickly) for $\sigma_n^{(1,2)}$ with increasing $n$, so truncating the system can be justified. For example,  $\sigma_0^{(\ell)} \sim 1/\ell^2$ from (\ref{eq:mean_singular_values}), 
and figure \ref{fig:sigma_subdom_scaling_PPF}(b) indicates that  $\sigma_n^{(1)} \sim 1/n$ for $n\gtrsim2$.

%
%

%
%
\begin{bottomfigure}
    \captionsetup{width=\columnwidth}
    \centering
    \includegraphics[width=0.75\linewidth]{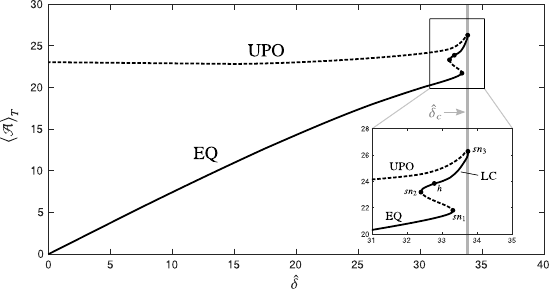}
    \caption{Bifurcation diagram for the reduced amplitude equations (\ref{eq:PPF_reduced_sys}), showing the slow-time averaged norm of the amplitudes $\mathcal{A} = \sqrt{\sum_{n,j}|A_n^{(j)}|^2}$ as a function of the forcing amplitude $\hat{\delta}$.  Solid/dotted lines denote stable/unstable solutions respectively.}
    \label{fig:A0_bif_PPF}
\end{bottomfigure}

\subsection{Results} \label{sec:PPF_asym_results}

The solutions to the reduced system of amplitude equations (\ref{eq:PPF_reduced_sys}) are investigated using a  truncation $\ell \in \{1,2\}$ and $n= \{0,\pm1,\pm2\}$, reducing the WNA system to just 2 real (for $n=0$) and 4 complex (for $n \in \{1,2\}$) degrees of freedom.  For simplicity, we set the forcing to have equal components in the two leading right singular vectors, i.e. $\boldsymbol{\hat{f}} = (\bv_1^{(1)} + \bv_1^{(2)})/\sqrt{2}$.  Other choices of forcing have also been tested, with the solutions of the amplitude equations maintaining a broadly similar qualitative behaviour.

\begin{figure}
    \centering
    \captionsetup{width=\columnwidth}
    \includegraphics[width=1\linewidth]{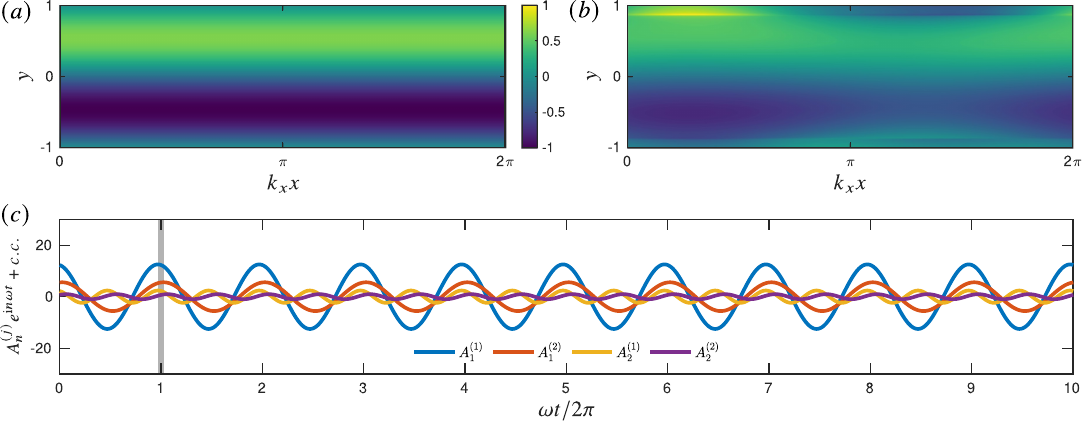}
    \caption{Stable solution on EQ branch at $\hat{\delta}=20$, reconstructed at $\Rey=10^6$, showing $(a)$ the streamwise component of the mean flow, and $(b)$ an instantaneous snapshot of the streamwise velocity at $\omega t/2\pi = 1$.  In both cases, we show the deviation from the laminar profile and the colourmap is normalised by the maximum velocity.
    $(c)$ The corresponding fast-time evolution of the oscillating modes.  The vertical grey line denotes the time of the snapshot shown in $(b)$.}
    \label{fig:flow_vis_20}
\end{figure}

The bifurcation diagram obtained for this system is depicted in figure \ref{fig:A0_bif_PPF}, which shows the slow-time average  of the amplitudes as a function of the applied forcing amplitude $\hat{\delta}$.  There is an equilibrium solution (EQ) that deviates from the laminar solution $U(y)$ as the forcing amplitude is increased from zero, the mean of which achieves a finite amplitude as a consequence of the stabilising nonlinear terms. The flow fields on EQ feature a strong mean shear associated with $\bu_0^{(2)}$ which increasingly dominates the other modes as $\hat{\delta}$ is increased (as shown for $\hat{\delta}=20$ in figure \ref{fig:flow_vis_20}$(a)$).  Despite the dominating mean shear, activity in the critical layers is still visible in instantaneous snapshots of the flow (figure \ref{fig:flow_vis_20}$(b)$), which oscillate periodically on the fast timescale.  The amplitude of these oscillations decreases with increasing $n$ (figure \ref{fig:flow_vis_20}$(c)$) as expected from the previous analysis of the singular values in figure \ref{fig:sigma_subdom_scaling_PPF}$(b)$. 

\begin{figure}
    \centering
    \captionsetup{width=\columnwidth}
    \includegraphics[width=1\linewidth]{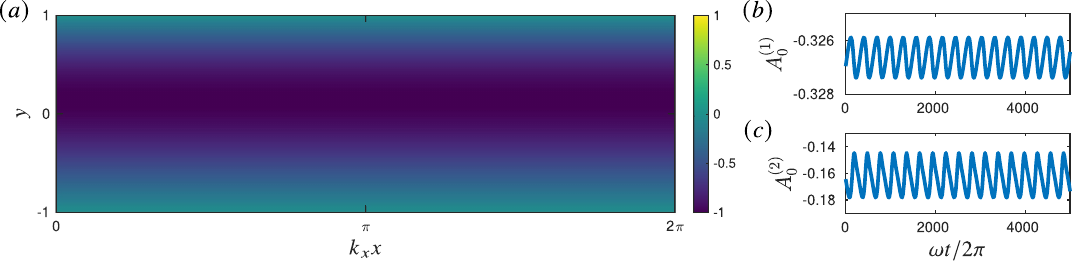}
    \caption{$(a)$ The streamwise component of the mean flow on the LC branch at $\hat{\delta}=33.4$ for $\Rey=10^6$.  Again, we show the deviation from the laminar profile and the colourmap is normalised by the maximum velocity.  The temporal evolution of the mean modes $A_0^{(1)}$ and $A_0^{(2)}$, which vary on the slow timescale, are shown in $(b,c)$ respectively.}
    \label{fig:flow_vis_33_4}
\end{figure}

Eventually as $\hat{\delta}$ is increased, the EQ solution collides in a saddle-node bifurcation at $\hat{\delta} \approx 33.25$ (marked as $sn_1$ in figure \ref{fig:A0_bif_PPF}).
The unstable branch connects to a second saddle-node bifurcation ($sn_2$) at $\hat{\delta} \approx 32.37$, giving rise to a new stable equilibrium.  This new stable equilibrium undergoes a supercritical Hopf bifurcation ($h$) at $\hat{\delta}\approx32.37$ leading to a stable limit cycle (LC) oscillating on the slow timescale.  Unlike EQ, solutions on LC are dominated by $\bu_0^{(1)}$, as observed in the mean flow shown in figure \ref{fig:flow_vis_33_4}$(a)$, which acts against the laminar profile.  The evolution of the mean amplitudes is depicted in figure \ref{fig:flow_vis_33_4}$(b,c)$, which shows a clear timescale separation from the forcing frequency.  Fast-time oscillations are still observed instantaneously in the critical layers and close to the walls, with the overall slow-time amplitude varying due to forcing from the mean amplitudes.
As the forcing is increased further, the limit cycle is eventually destroyed by colliding with an unstable periodic orbit (UPO) in a saddle-node bifurcation ($sn_3$) at $\hat{\delta}_c \approx 33.8$.  Above this forcing amplitude, no stable solutions have been found, and all initial conditions grow into a regime of different scaling associated with the fully nonlinear (presumably `turbulent') regime.   

For all values of $0 \leq \hat{\delta} \leq \hat{\delta}_c$ tested, we have been able to find initial conditions for which the transition to the fully nonlinear regime is triggered, for example, for $\hat{\delta} \leq \hat{\delta}_c$ choosing an initial condition with larger amplitude than the UPO branch will result in a transition to the fully nonlinear regime.  This suggests that the system of amplitude equations derived here can be used to investigate the conditions under which subcritical transitions are triggered as a function of the initial conditions, forcing profile, and forcing amplitude at high $\Rey$, at extremely reduced computational cost compared to fully nonlinear methods.

%
%
\subsection{Comparison to a previous approach} \label{sec:duc_compare}

The system of weakly nonlinear amplitude equations (\ref{eq:PPF_reduced_sys}) derived  in the previous section has a very different structure to that argued for in  \cite{ducimetiere2022weak}.
There, it was assumed that the leading singular value $\sigma_1^{(1)}$ is asymptotically separated from the remaining singular values of $\mathcal{R}(\mathrm{i}\omega)$ as $\eps \rightarrow 0$, so only $a_1^{(1)}$  appeared at leading order in the solution. Following classical WNA, an expansion of the form
\begin{align}
    \boldsymbol{q} &= \varepsilon^{1/2} A_1^{(1)}(T)\boldsymbol{u}_1^{(1)}(y) e^{\mathrm{i}(k_x x+\omega t)} 
    + \varepsilon\big[\hat{\boldsymbol{u}}_0(y,T) + \hat{\boldsymbol{u}}_2(y,T) e^{2\mathrm{i}(k_x x+\omega t)} \big] \nonumber \\
    & \hspace{7cm}+ \mathcal{O}(\varepsilon^{3/2}) +c.c.,
    \label{eq:duc_expansion}
\end{align}
was then pursued where $\hat{\bu}_0$ and $\hat{\bu}_2$ are general $\mathcal{O}(1)$ velocity fields (see equation (32) in \citet{Ducimetiere_PRE_2025}).
The key scaling $a_1^{(1)} = \eps^{1/2}A_1^{(1)}$ is necessitated by the fact that the leading nonlinear term in the $A_1^{(1)}$-amplitude equation then must be cubic in $A_1^{(1)}$ due to the quadratic nonlinearity of the Navier--Stokes equations. Then balancing the linear term with this nonlinear term and the forcing gives $a_1^{(1)}=\mathcal{O}(\sqrt{\eps})$ and $\delta=\mathcal{O}(\eps^{3/2})$ (\S II. B., pg. 10, \citet{Ducimetiere_PRE_2025}).  
As in normal WNA, to obtain the coefficient of this nonlinear term requires first solving the $\mathcal{O}(\eps)$ problems, 
\begin{subequations}
    \begin{align}
        n=0: \qquad\mathcal{R}^{-1}(0)\,\hat{\bu}_0 &= -2|A_1^{(1)}|^2 \,C(\bu_1^{(1)}, \bu_1^{(1)*}), \qquad\quad \label{eq:n0_duc}\\
        n=2: \hspace{0.4cm}\mathcal{R}^{-1}(2\mathrm{i}\omega)\,\hat{\bu}_2 &= -(A_1^{(1)})^2 \,C(\bu_1^{(1)}, \bu_1^{(1)}),
    \end{align}
\end{subequations}
where (using their notation) $C(\boldsymbol{a},\boldsymbol{b}):= \tfrac{1}{2}(\mathcal{N}(\boldsymbol{a},\boldsymbol{b}) + \mathcal{N}(\boldsymbol{b},\boldsymbol{a}))$.
The nonlinear interaction of these $O(\eps)$ fields with the $A_1^{(1)}$ mode then leads to the following amplitude equation:
\begin{equation}
    \Ov_1^{11}\partial_T A_1^{(1)} = -A_1^{(1)} - (c_0+c_2) A_1^{(1)}|A_1^{(1)}|^2 + \langle \bv_1^{(1)}, \boldsymbol{\hat{f}}\rangle \hat{\delta},
    \label{eq:duc_amp}
\end{equation}
where
\begin{subequations}
\label{eq:duc_c0_c2}
    \begin{alignat}{1}
        c_0 &:= \langle\boldsymbol{v}_1^{(1)}, 2C(\boldsymbol{u}_1^{(1)}, -2\mathcal{R}(0)C(\boldsymbol{u}_1^{(1)*}, \Bar{\boldsymbol{u}}_1^{(1)}))\rangle,  \\
        c_2 &:= \langle\boldsymbol{v}_1^{(1)}, 2C(\Bar{\boldsymbol{u}}_1^{(1)*}, -\mathcal{R}(2\mathrm{i}\omega)C(\boldsymbol{u}_1^{(1)},{\boldsymbol{u}}_1^{(1)}))\rangle.
    \end{alignat}
\end{subequations}
Here, $T=\eps t$ is a slow time and $\delta = \varepsilon^{3/2}\hat{\delta}$.

Our analysis indicates certain shortcomings in the derivation of this amplitude equation.
The first is that  $\sigma_1^{(2)} = \mathcal{O}(\sigma_1^{(1)})$ so $a_1^{(1)}$ is not the sole response at the forcing frequency. Secondly, $c_0$ and $c_2$ are not $\mathcal{O}(1)$ quantities.  In fact, the results of the previous section (\S\ref{sec:nonlinear_red_PPF}) indicate that $c_0, c_2 = \mathcal{O}(\eps^{-5/3})$ as $\eps \rightarrow 0$ due to the non-trivial nonlinear interaction of the singular vectors with each other, the application of the resolvents which do not have a norm of $\mathcal{O}(1)$, and the subsequent projection onto the right singular vectors.  This dependence on $\eps$ breaks the asymptotic consistency of the amplitude equation (\ref{eq:duc_amp}) to the extent that the mean response actually dominates the directly-forced fields, as shown here.

Nevertheless, their equation (2.12) still appears to capture the behaviour of the system (see figure 10 in \citet{ducimetiere2022weak}). Concentrating on their $\omega_0=0.3810$ example (\,figure 10(a)\,), the observed gain, their $G:=a_1^{(1)}/\delta$, displays three regimes as the forcing is increased. Initially, $G$ is constant  for very small forcing, which is the linear response. Then there is an  increase in $G$ to a maximum and decrease beyond this, which is presumably the WNA regime where all three terms in their equation (2.12) balance. Finally, there is a  regime of sustained decay where the forcing must be balanced by the nonlinear term only,  $\hat{\delta} \sim  A^3$, which in turn implies that $G = A/(\varepsilon\hat{\delta}) \sim \hat{\delta}^{-2/3}$.  This last regime comprises the bulk of the comparison in their figure 10(a) yet is formally beyond the weakly nonlinear regime. Additionally, \cite{ducimetiere2022weak} only consider $\Rey=3000$ and therefore one finite value of $\eps$, whereas the asymptotic regime arises for $\Rey\gtrsim10^5$ (e.g. see figure \ref{fig:sigma_scaling_PPF}). These considerations suggest examining the pre-asymptotic regime at $\Rey=3000$ to better understand their results.

%
%
\subsection{Pre-asymptotic regime} \label{sec:PPF_pre_asy}

In the pre-asymptotic regime, the maximum singular value $\sigma_1^{(1)} = \mathcal{O}(\Rey^{3/2})$ is obtained by varying $\omega = \mathcal{O}(\Rey^{-1/5})$ when $\Rey\lesssim 4\times10^5$ (see figure \ref{fig:sigma_scaling_PPF}). It is therefore tempting to use this local scaling to define $\varepsilon$ and to perform an asymptotic reduction based on the local scaling of the various terms.  Not surprisingly, however, it proves impossible to extract any clean scalings for the various terms due to the limited range of $\Rey$ and the finiteness of $\eps$ (further details are provided in appendix \ref{app:local_scaling}).

Nevertheless, there is a small set of singular values that numerically dominate the others at $\Rey=3000$, suggesting that it may be possible to truncate the equations under a low-rank assumption.  For $n=1$, the singular values decrease quickly with increasing $j$ (e.g. $\sigma_1^{(1)} \approx 14\, \sigma_1^{(2)}$), and also for $|n|\geq 2$ ($\sigma_1^{(1)} \gtrsim 16\, \sigma_n^{(1)}$).  However, the prefactor of the subdominant $n=0$ singular values decay more slowly with $j$ than for $n\neq0$ (e.g. $\sigma_0^{(1)} \approx 4\,\sigma_0^{(2)}$), and therefore, since $\sigma_0^{(1)} \approx 3\,\sigma_1^{(1)}$, the most important amplified modes will be associated with $\sigma_1^{(1)}$ and a collection of $\sigma_0^{(j)}$ terms. Considering the symmetry of the associated singular vector fields, the truncated equations that retain only $a_1^{(1)}$ and $a_0^{(j)}$ amplitudes are 
\begin{subequations}
\label{eq:PPF_truncate}
    \begin{alignat}{1}
        \Delta \Ov_1^{11}\partial_Ta_1^{(1)} &= - \frac{1}{\sigma_1^{(1)}}a_1^{(1)} + \sum_{\ell\,{\rm even}} 
        \big(\NL_{1,0}^{11\ell}+ \NL_{0,1}^{1\ell1}\big)a_1^{(1)}a_0^{(\ell)} + \langle \bv_n^{(1)}, \boldsymbol{\hat{f}}\rangle   \delta, \label{eq:PPF_truncate_a1}\\
        \Delta\partial_Ta_0^{(j)} &= - \frac{1}{\sigma_0^{(j)}}a_0^{(j)} + \big(\NL_{1,-1}^{j11} + \NL_{-1,1}^{j11}\big)|a_1^{(1)}|^2, \quad\mathrm{for}\: j\in \{1,3,5,\ldots\}, \label{eq:PPF_truncate_a0}
    \end{alignat}
\end{subequations}
where we note that $\NL_{1,-1}^{j11}=\NL_{-1,1}^{j11*}$.

In contrast to the asymptotic regime, antisymmetric mean modes ($j \in\{2,4,6,\ldots\}$) do {\it not} appear in the equations as they would require nonlinear coupling with $a_1^{(2)}$ to be excited due to their spatial symmetry.  However, $a_1^{(2)}$ has been omitted in the truncation since $\sigma_1^{(2)}$ is much smaller than $\sigma_1^{(1)}$ at this $\Rey$. 
More intriguingly, however, this system of equations is closely related to final equation (\ref{eq:duc_amp}) obtained by \citet{Ducimetiere_PRE_2025} if solutions are restricted to equilibria ($\partial_T=0$).  Then, equation (\ref{eq:PPF_truncate_a0}) can be substituted into equation (\ref{eq:PPF_truncate_a1}) to obtain a cubic equation for $a_1^{(1)}$.  
The only difference between the amplitude equations is that in (\ref{eq:PPF_truncate_a0}) modes which do not contribute due to symmetry constraints have been dropped as have the $n=2$ modes due to their smaller singular values (although they could be included in an equivalent way without changing the structure of the amplitude equation).  

The reconstructed solution from these equations, however,  is still different from that proposed by \cite{ducimetiere2022weak}, where it  was assumed that the leading-order term is $a_1^{(1)}$ with $a_0^{(j)}$ appearing at higher order (see expansion (\ref{eq:duc_expansion})).  In contrast,  here leading $a_0^{(j)}$ amplitudes are often comparable in magnitude to $a_1^{(1)}$ in the numerical solution of the amplitude equations (\ref{eq:PPF_truncate}).  This distinction is important since the results of the DNS, used in \citet{ducimetiere2022weak} to validate the solution, were filtered to only include the $n=1$ component and so were blind to the possibly large mean modes.  

%
%
\begin{figure}
    \captionsetup{width=\columnwidth}
    \centering
    \includegraphics[width=1\linewidth]{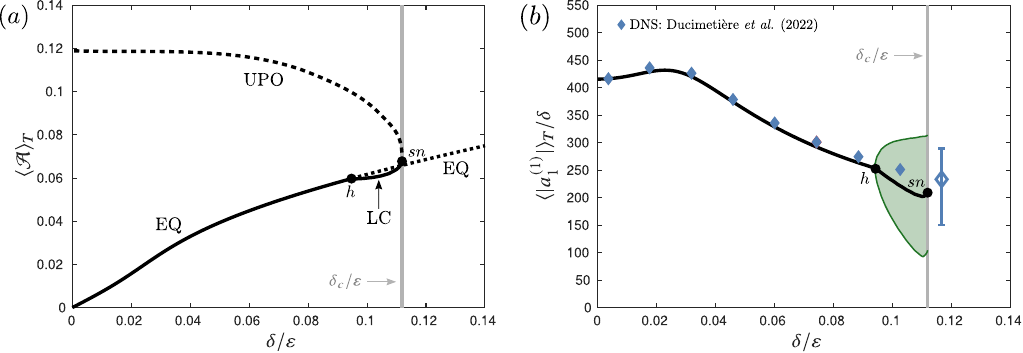}
    \caption{Bifurcation diagram for the reduced amplitude equations (\ref{eq:PPF_truncate}) at $\Rey=3000$, showing the slow-time averaged norm of $(a)$ all amplitudes $\mathcal{A} = \sqrt{|a_1^{(1)}|^2 + \sum_j|a_0^{(j)}|^2}$ in the truncated system, and $(b)$ the oscillatory amplitude $a_1^{(1)}$ normalised by the forcing amplitude $\delta$ (the harmonic gain), as a function of the forcing amplitude $\delta$.  Here, the forcing is normalised by $\varepsilon$ for easy comparison with \citet{ducimetiere2022weak}.  Solid/dotted lines denote stable/unstable solutions respectively.  The green shaded region denotes the standard deviation of the oscillations from the mean.}
    \label{fig:bif_PPF_3000}
\end{figure}

The truncated equations \eqref{eq:PPF_truncate} additionally capture time dependence on the slow-timescale (which was ignored in \citet{Ducimetiere_PRE_2025} or appeared at higher asymptotic order in \cite{ducimetiere2022weak}).  Balancing the linear part of the $n$\,=\,0 equations and the slow-time derivative gives $\Delta = \mathcal{O}(\Rey^{-1})$, which is thus consistently separated from the forcing timescale $\omega = \mathcal{O}(\Rey^{-1/5})$.  A similar timescale is obtained from the analysis of the $n$\,=\,1 equation (although the $\sigma_1^{(1)}$ scaling is only local) and thus, slow-time dependence is included in all components of the truncated equations (\ref{eq:PPF_truncate}).

\begin{figure}
    \captionsetup{width=\columnwidth}
    \centering
    \includegraphics[width=1\linewidth]{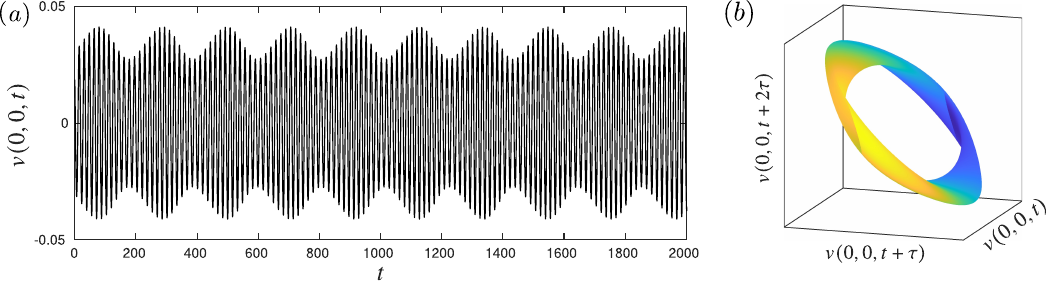}
    \caption{$(a)$ Temporal variation of the wall-normal velocity $v$ at $(x,y)=(0,0)$ for $\Rey=3000$ at $\delta/\varepsilon = 0.105$.  Since the wall-normal component of the mean ($n=0$) modes is zero, this component of the velocity only depends on the $a_1^{(1)}$ amplitude of the solution.  $(b)$ Attractor of the system constructed using time-delay embedding, with $\tau=20$.  The colour map corresponds to $v(0,0,t+3\tau)$.}
    \label{fig:v_PPF_3000}
\end{figure}

The truncated equations have been solved numerically for $\Rey=3000$, including the leading nine symmetric modes for $n$\,=\,0 thereby giving a system of ODEs with 1 complex  and 9 real degrees of freedom.  Bifurcation diagrams describing the solutions as a function of the forcing amplitude $\delta$ are shown in figure \ref{fig:bif_PPF_3000}.  In figure \ref{fig:bif_PPF_3000}$(a)$, the time-average of the norm of the vector of all amplitudes is shown, whereas in figure \ref{fig:bif_PPF_3000}$(b)$ only the norm of $a_1^{(1)}$ is shown to compare to the DNS results reported in \citet{ducimetiere2022weak}.  We see a qualitative structure similar to the solutions of the reduced system in the asymptotic regime (\S\ref{sec:PPF_asym_results}), where an equilibrium solution (EQ) grows from the laminar flow with increased $\delta$.  This equilibrium solution is equivalent to the solution obtained by \citet{ducimetiere2022weak} and agrees well with DNS (\ref{fig:bif_PPF_3000}$(b)$).  As $\delta$ is increased further this equilibrium undergoes a supercritical Hopf bifurcation at point $h$ generating a stable limit cycle solution (LC).  Since this limit cycle oscillates on the slow-timescale, the full reconstructed flow field, which additionally oscillates on the forcing timescale, exhibits quasiperiodic behaviour.  This can be seen in the wall normal velocity at the mid-plane, as shown in figure \ref{fig:v_PPF_3000}$(a)$.  The attractor is reconstructed using time-delay embedding (figure \ref{fig:v_PPF_3000}$(b)$), which suggests that this limit cycle in the amplitudes is a stable invariant 2-torus in the original variables $\boldsymbol{q}$.  This solution is not captured by the asymptotic approach of \citep{ducimetiere2022weak} but displays similar behaviour to the DNS at equivalent values of the forcing amplitude (compare figure \ref{fig:v_PPF_3000}$(a)$ with the inset of figure 10$(a)$ in \citet{ducimetiere2022weak}).  At $\delta_c/\varepsilon \approx 0.111$, the LC is destroyed in a saddle-node bifurcation with an unstable periodic orbit (UPO) at point $sn$. Above this threshold, 
similar to the results presented in \S\ref{sec:PPF_asym_results}, 
no stable solution to the truncated system is found.

To summarise, the promising results produced at low $\Rey$ by the amplitude equation written down in \citet{Ducimetiere_PRE_2025} are also  obtained by a truncated version of the equations, although these are in no way an asymptotically consistent set for $\eps \rightarrow 0$.

%
%

%
%
\section{A model ODE problem} \label{sec:model_prb}

Having applied an asymptotic reduction to plane Poiseuille flow, we now examine the approach in greater detail using the simplest possible setting of a two-dimensional system of nonlinear ODEs.  Here, all the various coefficients needed to asymptotically reduce the problem (equation (\ref{eq:coefficients})) and their dependence on $\varepsilon$ can be obtained analytically.  This reduced complexity provides a convenient setting in which to explore the following fundamental questions: (i) under what conditions can the method be applied, for example, is $\varepsilon$ always well-defined?; and (ii) to what extent are the amplitude equations derived via this approach universal?

%
%
\subsection{Problem definition and initial analysis} \label{sec:model_prb_def}

The linear stability of a streamwise, viscous parallel shear flow $U(y)$ is governed by the Orr--Sommerfeld--Squire system, which expresses the evolution of the wall-normal velocity and wall-normal vorticity perturbations, $\tilde{v}(x,y,z,t) = \hat{v}(y,t)\exp[\mathrm{i}(\alpha x +\beta z)]$ and $\tilde{\eta}(x,y,z,t) = \hat{\eta}(y,t)\exp[\mathrm{i}(\alpha x +\beta z)]$, respectively, as
\begin{equation}
    \frac{\partial}{\partial t}
    \begin{pmatrix}
        k^2-\partial_y^2 & 0 \\ 0 & 1
    \end{pmatrix}
    \begin{pmatrix}
        \hat{v} \\ \hat{\eta}
    \end{pmatrix}
    =
    \begin{pmatrix}
        -\mathcal{L}_{OS} & 0 \\ -\mathrm{i}\beta\partial_y U & -\mathcal{L}_{SQ}
    \end{pmatrix}
    \begin{pmatrix}
        \hat{v} \\ \hat{\eta}
    \end{pmatrix},
\end{equation}
where $k^2 := \alpha^2+\beta^2$, 
\beq
\mathcal{L}_{OS} := \mathrm{i}\alpha U(k^2-\partial_y^2)+\mathrm{i}\alpha\partial_y^2U + \Rey^{-1}(k^2-\partial_y^2)^2
\eeq
is the Orr-Sommerfeld operator and 
\beq
\mathcal{L}_{SQ} := \mathrm{i}\alpha U + \Rey^{-1}(k^2-\partial_y^2)
\eeq
is the Squire operator.  For streamwise independent disturbances ($\alpha=0$) and assuming that $k^2-\partial_y^2 =\mathcal{O}(1)$, $\mathcal{L}_{OS} = \mathcal{O}(\Rey^{-1})$ and $\mathcal{L}_{SQ} = \mathcal{O}(\Rey^{-1})$.  This motivates the following two-dimensional, harmonically-forced ODE system, which evolves the state $\boldsymbol{q} = (v(t)\,\,\eta(t) )^T$ according to
\begin{equation}
    \frac{\mathrm{d}}{\mathrm{d} t}
    \begin{pmatrix}
        v \\
        \eta
    \end{pmatrix}
     = 
     \underbrace{
     \begin{pmatrix}
         -\Rey^{-1} & 0 \\
         1 & -b\Rey^{-1}
     \end{pmatrix}
     \begin{pmatrix}
        v \\
        \eta
    \end{pmatrix}
    }_{\mathcal{L}\boldsymbol{q}}
    + 
    \underbrace{
    \begin{pmatrix}
        \eta^2 \\
        -v\eta
    \end{pmatrix}
    }_{\mathcal{N}(\boldsymbol{q},\boldsymbol{q})}
    + \:\delta({\boldsymbol{\hat{f}}}e^{\mathrm{i}\omega t} + {c.c.}),
    \label{eq:model-sys}
\end{equation}
where $b$ is an $\mathcal{O}(1)$ constant and  $\boldsymbol{\hat{f}} = -\mathrm{i}(1/2\,\, 1/2)^T$  so the forcing is $\sin(\omega t)$ in both components. The linear term is supplemented with a quadratic energy-preserving nonlinearity, i.e. $\bq \cdot \mathcal{N}(\bq,\bq)=0$, consistent with the Navier--Stokes equation \citep[e.g. see also][]{Trefethen93,baggett1997low,liu2020input}. 

\begin{figure}
    \captionsetup{width=\columnwidth}
    \centering
    \vspace{5pt}
    \includegraphics[width=0.85\linewidth]{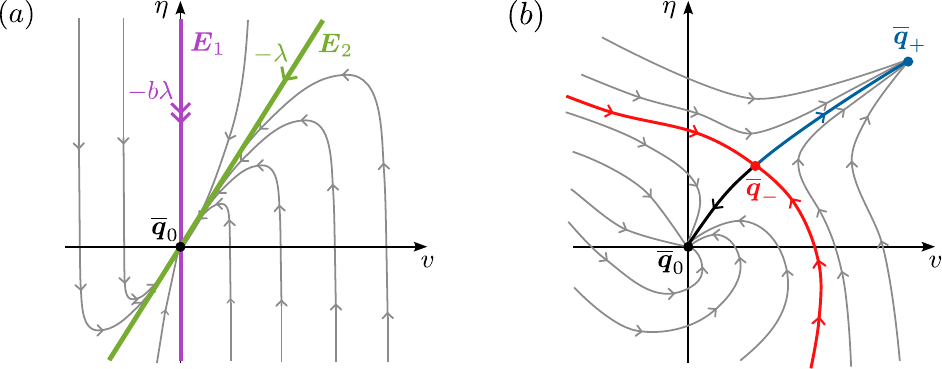}
    \caption{Illustrative sketch of the phase portraits with $b>1$ for $(a)$ the unforced linearised system showing the two stable eigendirections $\boldsymbol{E}_1$ (purple) and $\boldsymbol{E}_2$ (green), and $(b)$ the unforced full nonlinear system showing the separatrix (red) defining the basin of attraction for the two stable equilibria $\overline{\boldsymbol{q}}_{0}$ and $\overline{\boldsymbol{q}}_{+}$, and the heteroclinic orbits between the fixed points (black/blue).}
    \label{fig:phase_port_sketch}
\end{figure}

The linear operator $\mathcal{L}$ is non-normal and linearly stable  with eigenvalues $-b/\Rey$ and $-1/\Rey$.  Non-normality manifests itself by the corresponding eigenvectors $\boldsymbol{E}_1 = (0 \,\,1)^T$ and $\boldsymbol{E}_2 = (1 \,\, \Rey/(b-1))^T$ becoming more aligned as $\Rey\rightarrow \infty$, giving rise to large transient growth.  When $b\neq1$, the solution takes the general form
\begin{equation}
    \boldsymbol{q}(t) = c_1\exp({-b t/\Rey})\boldsymbol{E}_1 + c_2\exp({-t/\Rey})\boldsymbol{E}_2
\end{equation}
whereas for $b=1$, $\opL$ is defective and the  generalised eigenvector $(1 \,\,t)^T$ replaces $\boldsymbol{E}_2$.
The transient growth that is possible can be observed in the sketch of the phase portrait of the linearised system in figure \ref{fig:phase_port_sketch}$(a)$, where the Euclidean norm of many trajectories grows initially before being attracted back to the stable equilibrium $\overline{\bq}_0={\bf 0}$.  The timescale for the system to relax back to its equilibria is $\mathcal{O}(\Rey)$.

%
%

The unforced nonlinear system, i.e. (\ref{eq:model-sys}) with $\delta=0$, admits three equilibrium solutions
\begin{equation}
\overline{\boldsymbol{q}}_0 = 
\begin{pmatrix}
    0 \\
    0
\end{pmatrix},
\quad
    \overline{\boldsymbol{q}}_{\pm}
    = 
    \left(\begin{array}{c}
        \,\,\,\,\Rey \Big[1\pm \sqrt{1-4b/\Rey^2}\,\Big]^2/4\,\, \\
        \Big[1 \pm \sqrt{1-4b/\Rey^2}\,\Big]/2
    \end{array}\right).
    \label{eq:unforced_eq}
\end{equation}
For the range of $\Rey$ considered below, the equilibrium state $\overline{\boldsymbol{q}}_{+}$ is a stable equilibrium point, and the equilibrium $\overline{\boldsymbol{q}}_{-}$ is a saddle point whose stable manifold defines a separatrix between the two stable equilibria (shown in red along with blue and black heteroclinic orbits in figure \ref{fig:phase_port_sketch}$(b)$).

%
%
\subsection{Singular values and asymptotic limits} \label{sec:singular_val}

To establish an asymptotic limit, we need to examine the behaviour of $\varepsilon\equiv1/\sigma_1^{(1)}$ as a function of the forcing frequency $\omega$ in the limit as $\Rey\rightarrow\infty$. Since $\mathcal{R}(\mathrm{i}n\omega) \in \mathbb{C}^{2\times 2}$, the singular values can be derived  analytically and are found to depend strongly on $\omega$.  Writing $\omega = \mathcal{O}(\Rey^{-\alpha})$, then 
\begin{equation}
\sigma_1^{(1)} =
    \begin{cases}
        \: \mathcal{O}(1/\Rey^{|\alpha|}), & \alpha<0, \\
        \: \mathcal{O}(1), & \alpha = 0,\\
        \: \mathcal{O}(\Rey^{2\alpha}), & 0 <\alpha< 1, \\
        \: \mathcal{O}(\Rey^{2}), & \alpha \geq 1
    \end{cases}
    \quad,
    \qquad
    \sigma_1^{(2)} =
    \begin{cases}
        \: \mathcal{O}(1/\Rey^{|\alpha|}), & \alpha<0, \\
        \: \mathcal{O}(1), & \alpha \geq 0.
    \end{cases}
    \label{eq:singular_val_scale_cases}
\end{equation}
(See appendix \ref{app:svd_model} for details.) If $\alpha\leq 0$, the maximum gain of the system is bounded from above and so $\varepsilon  \not \rightarrow 0$ as $Re \rightarrow \infty$. The transient growth mechanism simply does not have the time to act if the forcing frequency is sufficiently large. In contrast, if the forcing frequency is reduced as $Re$ is increased such that $\alpha>0$, then the maximum gain $\sigma_1^{(1)}\rightarrow\infty$ as $\Rey\rightarrow\infty$, and the  asymptotic limit $\varepsilon \rightarrow 0$ is well-defined.  The rate at which $\sigma_1^{(1)}$ diverges as $\Rey \rightarrow\infty$ also saturates at $\sim \Rey^2$ when the forcing timescale is comparable to or larger than the relaxation time.
Significantly, when $\alpha > 0$, the second singular value $\sigma_1^{(2)}=\mathcal{O}(1)$ as $\Rey\rightarrow\infty$, meaning the leading singular value is asymptotically separated from the other singular value in this regime.  

There are therefore two regimes where an asymptotic reduction is possible: (i) $0 <\alpha< 1$ and (ii) $\alpha \geq 1$.  For brevity, we shall now detail the asymptotic reduction of the system using the method described in \S \ref{sec:weak-nonlin-exp} for regime (i), leaving details of the reduction for regime (ii) to appendix \ref{app:red_alpha_1}.

%
%

\subsection{Asymptotic reduction when $0<\alpha<1$} \label{sec:red_alpha_0_1}

The general method outlined in \S\ref{sec:gen_system} requires an assessment of  the asymptotic scaling of the various coefficients (see \ref{eq:coefficients})  which weight the contribution of the various terms in equation (\ref{eq:final_gen_eq}).  The leading singular values are found to be
\begin{equation}
    {\sigma}_n^{(1)} =
    \begin{cases}
        \mathcal{O}(\varepsilon^{-1}), & n\neq 0 \\
        \mathcal{O}(\varepsilon^{-1/\alpha}), & n=0
    \end{cases}
\end{equation}
the left and right singular vectors satisfy
\begin{equation}
    \Ov_n^{11} =  \langle\boldsymbol{v}_n^{(1)},\boldsymbol{u}_n^{(1)}\rangle =
    \begin{cases}
        \mathcal{O}(\varepsilon^{1/2\alpha}+i\varepsilon^{1/2}), & n\neq 0 \\
        \mathcal{O}(\varepsilon^{1/2\alpha}), & n=0,
    \end{cases}
\label{overlap_11}
\end{equation}
\begin{equation}
    \Ov_n^{12} = \langle\boldsymbol{v}_n^{(1)},\boldsymbol{u}_n^{(2)}\rangle =
    \begin{cases}
        \mathcal{O}(\varepsilon^{1/2\alpha - 1/2}+i), & n\neq 0 \\
        \mathcal{O}(1), & n=0,
    \end{cases}
\label{overlap_12}    
\end{equation}
and 
\begin{equation}
\langle{\boldsymbol{v}_1^{(1)},\boldsymbol{\hat{f}}}\rangle = \mathcal{O}(1).
\end{equation}
These results are derived analytically in appendix \ref{app:svd_model} but for a dominant singular value, as here, it is expected that
\beq
 \Ov_n^{11} =  \langle\boldsymbol{v}_n^{(1)},\boldsymbol{u}_n^{(1)}\rangle = \frac{\rho(\mathcal{R})}{\sigma_1} \leq 1
\eeq
where $\rho(\mathcal{R})$ is the spectral radius of $\mathcal{R}$ and $\sigma_1$ the largest singular value of $\mathcal{R}$: see  appendix \ref{Gelfand}. For a normal matrix this is precisely 1 as the spectral radius, or equivalently the modulus of the largest eigenvalue, is equal to the largest singular value but for a non-normal matrix it is strictly less than one with the amount of inequality measuring the degree of  non-normality.

Not all projections can be small with, for example, the projection of $\bv_n^{(1)}$ onto the subdominant response vector $\boldsymbol{u}_n^{(2)}$ asymptotically larger than the projection onto dominant response vector $\boldsymbol{u}_n^{(1)}$.  Despite this, analysis of the linearised system indicates that $a_n^{(2)}$ is smaller than $a_n^{(1)}$ by a factor of $\varepsilon$ for $n\neq0$ and $\varepsilon^{1/\alpha}$ for $n=0$, and thus, the dominant (slow) time derivative term remains $a_n^{(1)}\langle\boldsymbol{v}_n^{(1)},\boldsymbol{u}_n^{(1)}\rangle$ for all $n$. The nonlinear 
interaction between the leading singular vectors is seen to be asymptotically larger than the other interactions and scales as 
\begin{equation}
    \NL_{n,m}^{111} = 
    \begin{cases}
        \mathcal{O}(1), & n=m=0,\\
        \mathcal{O}(1 + i\varepsilon^{1/2\alpha+1/2}), & n+m = 0, \\
        \mathcal{O}(1 + i\varepsilon^{1/2\alpha-1/2}), & \rm{otherwise}.
    \end{cases}
    \quad 
\end{equation}
This results in various cases for the scaling of terms that can appear at leading order in equation (\ref{eq:final_gen_eq}) depending on the value of $n$, as follows:
\begin{subequations}
\begin{alignat}{2}
    &n=0: \qquad\qquad &&(\ldots)\Delta  \partial_T a_0^{(1)} \varepsilon^{1/2\alpha} + (\ldots)\varepsilon^{1/\alpha} a_0^{(1)} = \mathcal{O}\big(\sum_m a_{-m}^{(1)}a_m^{(1)}\big), \qquad \label{eq:balance_n0}\\
    &|n|=1: \qquad\qquad &&(\ldots)\Delta  \partial_T a_n^{(1)} \varepsilon^{1/2} + \varepsilon a_n^{(1)} - \mathcal{O} \big( \sum_m a_{n-m}^{(1)}a_m^{(1)} \big) = \mathcal{O}(\delta ), \qquad \label{eq:balance_n1}\\
    &|n|\geq 2: \qquad\qquad && (\ldots)\Delta  \partial_T a_n^{(1)} \varepsilon^{1/2} + (\ldots)\varepsilon a_n^{(1)}  = \mathcal{O}\big(\sum_m a_{n-m}^{(1)}a_m^{(1)}\big). \qquad  \label{eq:balance_n2}
\end{alignat}
\end{subequations}

%
%
\subsubsection{Scalings} 

Balancing the linear and forcing terms in (\ref{eq:balance_n1}) gives $a_1^{(1)} \sim \delta/\varepsilon$. Since the $n=0$ mode is more strongly amplified than the $n \ne 0$ modes for $0<\alpha<1$, a large mean response is expected; accordingly, we also balance
the nonlinear self-interaction of the mean mode with the linear term in (\ref{eq:balance_n0}), which requires $a_0^{(1)} \sim \varepsilon^{1/\alpha}$.  
Further recognising that a mean response can be directly excited only via the nonlinear self-interaction of the harmonically-forced mode, we insist that
$a_0^{(1)}a_0^{(1)} \sim a_{-1}^{(1)}a_1^{(1)} \sim \delta^2/\varepsilon^2$, yielding $\delta \sim \varepsilon^{1+ 1/\alpha}$ and $a_1^{(1)} \sim \varepsilon^{1/\alpha}$.  The slow time scaling arises from balancing $\Delta  a_0^{(1)} \varepsilon^{1/2\alpha}$ with $\varepsilon^{1/\alpha}a_0^{(1)}$ in the mean equation, allowing the prescription $\Delta = \varepsilon^{1/2\alpha} \sim\Rey^{-1} \ll \omega \sim \Rey^{-\alpha} \sim \varepsilon^{1/2}$ as required to be `slow'.  This slow timescale  is precisely the timescale associated with the relaxation of the linearised system to its equilibrium (\S\ref{sec:model_prb_def}) or equivalently that over which the most transient growth occurs.

Turning attention to $|n|>1$ amplitudes, the $n=2$ equation (\ref{eq:balance_n2}) implies $\varepsilon a_2^{(1)} \sim (a_1^{(1)})^2 \sim \varepsilon^{2/\alpha}$, and thus $a_2^{(1)} \sim \varepsilon^{(2/\alpha) - 1}$ 
so that the time derivative is asymptotically smaller. 
Considering higher values of $n$ (equation (\ref{eq:balance_n2})) similarly, so $\varepsilon a_n^{(1)} \sim a_{n-1}^{(1)}a_1^{(1)}$ then $a_n^{(1)} \sim a_{n-1}^{(1)} \varepsilon^{(1/\alpha) - 1}$ and so $a_n^{(1)}\ll a_{n-1}^{(1)}$ for $n>1$ (since $1/\alpha-1>0$).

In conclusion, for $\omega = \mathcal{O}(\Rey^{-\alpha})$ with $0<\alpha<1$, the only asymptotically consistent choice for each $a_n^{(1)}$, $\Delta$ and $\delta$ is:
\begin{my_indent}
    \item The leading amplitudes are $a_0^{(1)}\sim a_{1}^{(1)} = \mathcal{O}(\varepsilon^{1/\alpha})$;
    \item The amplitudes for $|n|>1$ are asymptotically smaller with $a_{1}^{(1)} \gg a_{2}^{(1)} \gg a_{3}^{(1)} \gg \cdots$ according to the relation $a_{n}^{(1)}\sim a_{n-1}^{(1)}\varepsilon^{(1/\alpha) - 1}$ for $n>1$;
    \item The forcing amplitude  $\delta = \mathcal{O}(\varepsilon^{1+ 1/\alpha})$;
    \item The slow time-scale $\Delta := \varepsilon^{1/2\alpha} \ll \omega = \mathcal{O}(\varepsilon^{1/2})$ is consistent.
\end{my_indent}

With these scalings, we now proceed to reduce and rescale the amplitude equations. For $n$\,=\,1, the linear term and the forcing terms dominate the nonlinear and slow-time derivative terms, enabling the amplitude $a_1^{(1)}$ to be solved for directly as
\begin{equation}
    a_1^{(1)} = \sigma_1^{(1)} \delta \langle \bv_n^{(1)}, \boldsymbol{\hat{f}}\rangle .
    \label{eq:soln_a1_scalar}
\end{equation}
For the $n$\,=\,0 equation, the linear, time derivative and nonlinear interactions involving $a_0^{(1)}$ and $a_1^{(1)}$ are all important to leading order as $\varepsilon\rightarrow0$.  Substituting the expression (\ref{eq:soln_a1_scalar}) 
for $a_{1}^{(1)}$ (\,=$a_{-1}^{(1)*}$\,) yields
\begin{equation}
    \Delta \Ov_0^{11}\partial_T a_0^{(1)}  + \frac{1}{\sigma_0^{(1)}} a_0^{(1)} = \NL_{0,0}^{111} \big(a_0^{(1)}\big)^2 + \big(\NL_{1,-1}^{111}+\NL_{-1,1}^{111}\big)\big|a_1^{(1)}\big|\,^2.
    \label{eq:a0_equation_model}
\end{equation}
Defining the rescaled amplitudes $a_0^{(1)} = \varepsilon^{1/\alpha}A_0^{(1)}$ and $a_1^{(1)} = \varepsilon^{1/\alpha}A_1^{(1)}$ so that $A_0^{(1)}$ and $A_1^{(1)}$ are $\mathcal{O}(1)$ amplitudes, (\ref{eq:a0_equation_model}) can be compactly expressed as
\begin{equation}
    \frac{\mathrm{d} A_0^{(1)}}{\mathrm{d}T} = -d_l A_0^{(1)} + d_{n} \big(A_0^{(1)}\big)^2 + d_f,
    \label{eq:amp_eqn_scalar}
\end{equation}
where the $\mathcal{O}(1)$ linear, nonlinear and forcing coefficients ($d_l$, $d_n$ and $d_f$ respectively) are defined as
\begin{equation}
    d_l:= \frac{1}{\varepsilon^{1/2\alpha}\,\sigma_0^{(1)}\Ov_0^{11}}, \quad
    d_n := \frac{\varepsilon^{1/2\alpha}\NL_{0,0}^{111} }{\Ov_0^{11}}, \quad
    d_f := \frac{\varepsilon^{1/2\alpha}\big(\,\NL_{1,-1}^{111}+\NL_{-1,1}^{111}\,\big)\big|A_1^{(1)}\big|\,^2}{\Ov_0^{11}},
    \label{eq:amp_scalar_coeff}
\end{equation}
where $A_1^{(1)} = \eps^{-(1/\alpha)-1} \delta \langle \bv_n^{(1)}, \boldsymbol{\hat{f}}\rangle $ is an $\mathcal{O}(1)$ constant amplitude directly measuring the forcing given (\ref{eq:soln_a1_scalar}).  Similarly, the original forcing amplitude is given as $\delta = \varepsilon^{1+ 1/\alpha}\hat{\delta}$.
As anticipated, the coefficient $d_l>0$, but the sign of $d_n$ and $d_f$ depend on the arbitrary sign choice of the $\bu_0^{(1)}$, $\bv_0^{(1)}$ pair.  For simplicity, we choose the sign of these singular vectors such that the forcing coefficient $d_f\geq0$ (giving $d_n>0$).

Once the amplitude equation \eqref{eq:amp_eqn_scalar} is solved, the asymptotic solution is reconstructed as 
\begin{equation}
    \boldsymbol{q} =
    \begin{pmatrix}
        v(t) \\
        \eta(t)
    \end{pmatrix}
    = \varepsilon^{1/\alpha}\Big[ A_0^{(1)}(\eps^{1/2\alpha} \, t) \boldsymbol{u}_0^{(1)} + A_1^{(1)} e^{\mathrm{i}\omega t}\boldsymbol{u}_1^{(1)} + c.c.\Big] + \mathcal{O}(\varepsilon^{(2/\alpha)-1}).
    \label{eq:soln_recon_w_gen}
\end{equation}
\vspace{0.5cm}

%
%
\begin{figure}
    \captionsetup{width=\columnwidth}
    \centering
    \vspace{10pt}
    \includegraphics[width=0.55\linewidth]{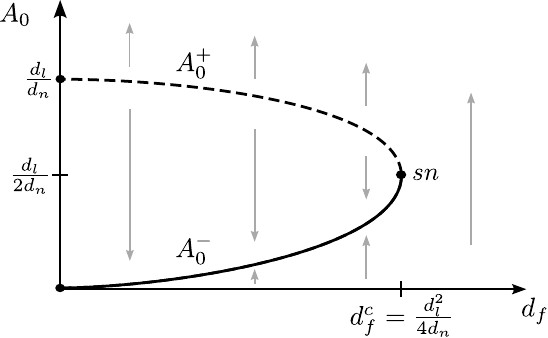}
    \caption{Sketch of bifurcation diagram for the amplitude equation (\ref{eq:amp_eqn_scalar}) showing the saddle-node bifurcation occurring at a critical value of the forcing $d_f^c$.  Solid lines denote stable solutions, and dashed lines denote unstable solutions.}
    \label{fig:bif_diagram_sketch}
\end{figure}

%
%
\subsubsection{Solution to the $0<\alpha<1$ amplitude equation} \label{sec:solutions_alpha_0_1}

The amplitude equation  (\ref{eq:amp_eqn_scalar}) can be integrated directly by separating variables, yielding
\beq
A_0(T) = \frac{A_0^-(\, A_0^+-A_0(0)\,)+A_0^+(\, A_0(0)-A_0^- \,)\exp( -(A_0^+-A_0^-)d_n t)  }
              {    \quad \, (\, A_0^+-A_0(0)\,)+    \quad \,\, (\,A_0(0)-A_0^-\,)\exp( -(A_0^+-A_0^-)d_n t)    }
\eeq
provided $A_0(0) \leq  A_0^+$ and $d_f < d_f^c$, where the pair of equilibria 
\beq
    {A}_0^{\pm} = \frac{d_l\pm \sqrt{d_l^2-4 d_n d_f}}{2d_n}
\label{equilibria}
\eeq
exist if  $d_f \leq d_f^c:=d_l^2/4d_n $ (and dropping $^{(1)}$ for clarity).
But it is more informative simply to rewrite (\ref{eq:amp_eqn_scalar}) as 
\begin{equation}
    \frac{\mathrm{d} A_0}{\mathrm{d}T} = d_n \big( A_0-A_0^+ \big) \big( A_0-A_0^-\big),
    \label{simple}
\end{equation}
and then the dynamics as a function of the initial condition $A_0(0)$ are clearly
\begin{equation}
    A_0(T) \rightarrow 
    \begin{cases}
        A_0^-,   & A_0(0)<A_0^+, \\
        \infty,  & A_0(0)>A_0^+,
    \end{cases}
    \qquad {\rm as}\,\, T\rightarrow \infty,
\end{equation}
for $d_f < d_f^c$.  At the critical forcing amplitude $d_f=d_f^c$, $A(T) \rightarrow A_0^-=A_0^+$ if $A_0(0) \leq  A_0^-$ and otherwise $A_0(T) \rightarrow \infty$ as $T \rightarrow \infty$. Finally, for  $d_f >  d_f^c$, $A_0(T) \rightarrow \infty$ for all initial conditions.  This behaviour is summarised in the sketch of the bifurcation diagram, shown in figure \ref{fig:bif_diagram_sketch}. 

%
%
\begin{figure}
    \captionsetup{width=\columnwidth}
    \centering
    \includegraphics[width=0.95\linewidth]{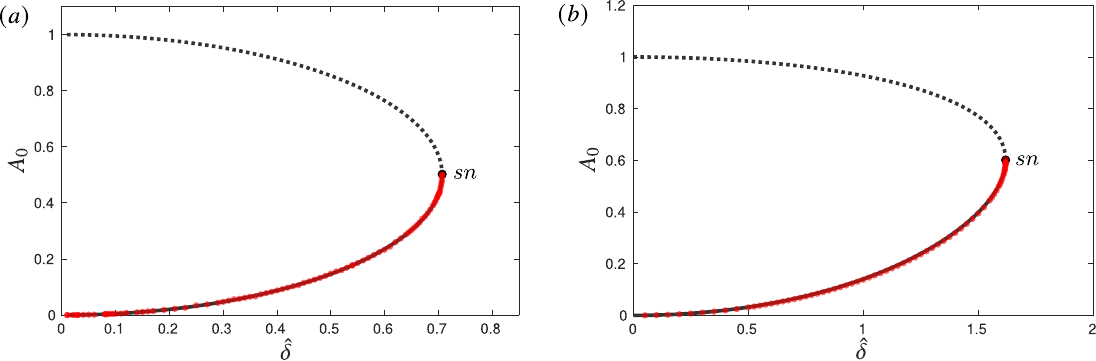}
    \caption{Bifurcation diagrams as a function of the rescaled forcing amplitude $\hat{\delta}$ computed from the reduced amplitude equations (black), showing stable (solid) and unstable/saddle (dotted) branches.  The black marker shows the saddle-node bifurcation point.  Red markers denote stable numerical solutions computed from the original fully nonlinear system (Eq. (\ref{eq:model-sys})) for $\Rey=10^{4}$, $b=1$ in $(a)$ the $0<\alpha<1$ regime (Eq. (\ref{eq:amp_eqn_scalar})) with $\alpha=1/2$ ($\omega = \Rey^{-1/2}$), and $(b)$ the $\alpha\geq1$ regime (Eq. (\ref{eq:amplitude_sys})) with $\alpha=1$ ($\omega=1/\Rey$).}
    \label{fig:bif_mean_compare_all}
\end{figure}

Therefore, in  the  weakly nonlinear regime, $0 < d_f< d_f^c$, there is a stable branch of solutions (the solid black line representing $A_0^-$ in figure \ref{fig:bif_mean_compare_all}$(a)$) which increase in amplitude up to a forcing threshold defined by the saddle node bifurcation at $d_f^c$.  In the original forcing amplitude, this threshold depends on $\Rey$ as
\beq
\delta_c = \mathcal{O}( \eps^{1+1/\alpha})=\mathcal{O}(Re^{-2(1+\alpha)}).
\eeq
Along with this stable branch, an unstable branch of solutions exists (dotted black line representing $A_0^+$ in figure \ref{fig:bif_mean_compare_all}$(a)$), with the associated amplitude decreasing as the forcing is increased.
If the undisturbed state is perturbed with sufficient amplitude to push the initial condition above the unstable branch, the solution of the amplitude equation transitions into a fully nonlinear regime characterised by a different scaling.  
Hence, the analysis has predicted both the upper forcing threshold $\delta_c$ ($d_f^c$) above which all initial conditions transition to the fully nonlinear state 
{\em and} also, for any $\delta<\delta_c$, identified the finite-amplitude initial conditions for which a transition is observed.
In both cases, the transition leads to system dynamics that are fully nonlinear, which, in the context of the original nonlinear phase portrait sketched in figure \ref{fig:phase_port_sketch}$(b)$, corresponds to the system approaching the fully nonlinear solution $\overline{\boldsymbol{q}}_+$.

\begin{figure}
    \captionsetup{width=\columnwidth}
    \centering
    \includegraphics[width=0.95\linewidth]{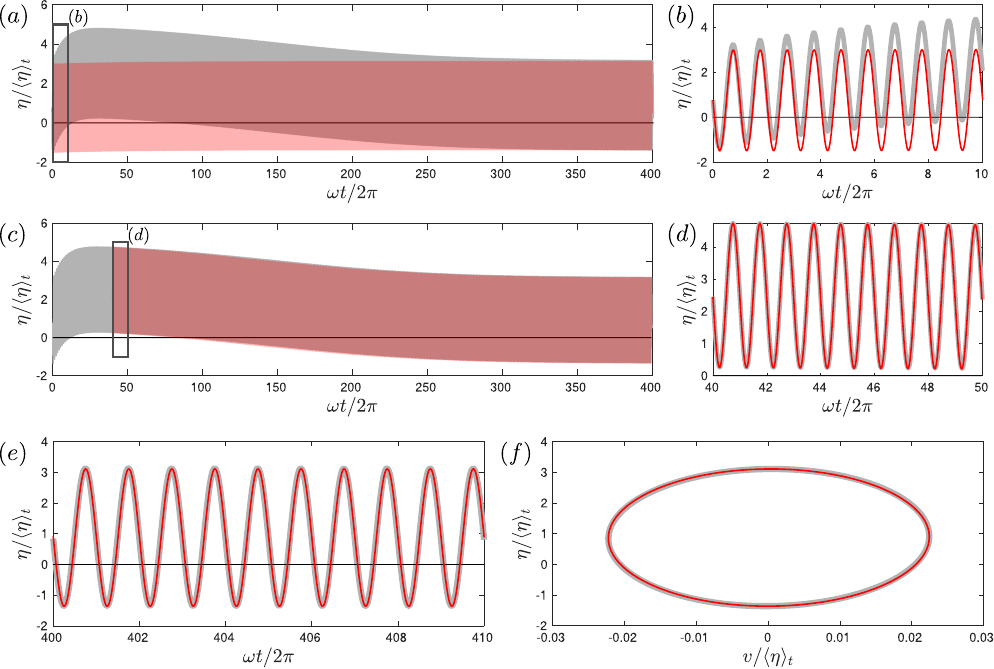}
    \caption{Comparison of the solutions of the fully nonlinear system (grey) and the amplitude equation (Eq. (\ref{eq:amp_eqn_scalar}), red) for two different initial conditions with $\Rey = 10^{4}$, $\omega = \Rey^{-1/2}$, $b=1$, $\hat{\delta} = 0.6$.  For $(a,b)$ the initial condition is taken from a simulation at lower $\hat{\delta}$ and for $(c,d)$ the initial condition for the amplitude equation is adjusted to match the fully nonlinear solution at $\omega t/2\pi=40$.  The applied harmonic forcing is shown as a function of time by the solid black lines.  The long-time behaviour is shown in both cases for $(e)$ the timeseries and $(f)$ the phase portrait of the solution.  In all cases, the components of the solution $(v,\eta)$ are normalised by the time average of $\eta$, denoted $\langle\eta\rangle_t$. \label{fig12}}
\end{figure}

To test the predictions, we first compare the attractor of the  fully nonlinear system (equation (\ref{eq:model-sys})) at $\Rey=10^{4}$ and $b=1$ for a range of forcing values below the critical threshold $\delta^c_f$. The amplitudes of the mean (over one forcing period) match very well with those of the stable $A_0^-$ branch from the amplitude equation: see figure \ref{fig:bif_mean_compare_all}(a).  To quantify this agreement,  for example, the amplitude equation (\ref{eq:amp_eqn_scalar}) predicts 
$\hat{\delta}_c \approx 0.707072$ when $\alpha=1/2$ whereas the full nonlinear system yields  $\hat{\delta}_c \approx 0.70711$. A comparison of the full and reduced systems for arbitrary initial conditions is a bit more subtle due to the assumption of slow time dynamics in the latter. In figures \ref{fig12}$(a,b)$, the evolution of the solution of the amplitude equation (\ref{eq:amp_eqn_scalar}) and the solution from the full nonlinear system (\ref{eq:model-sys}) is compared over time with both initiated at the origin.  In the initial stages of the evolution (figure \ref{fig12}$(b)$), the mean of both solutions grows in time.  However, the fully nonlinear solution (grey) grows more rapidly than the solution of the amplitude equation (red), reaching a maximal value at approximately 30 forcing periods before slowly decaying to a stable periodic solution (see the grey solution envelope in fig. \ref{fig12}$(a)$). This initial growth on the fast time scale falls outside of the scope of asymptotic reduction, whereas the slow decay is captured if the amplitude equation is reinitialised by matching the full solution at $\omega t/2 \pi=40$: see figure \ref{fig12}$(c$--$f)$.  In all cases, the large amplification from the non-normality is apparent from the comparative size of the response and the harmonic forcing (the latter is shown by a solid black line in figures \ref{fig12}$(a$--$e)$, which is close to zero since $\sigma_1^{(1)} = \mathcal{O}(\Rey)$ for $\alpha=1/2$). 

%
%

%
%
\subsection{Asymptotic reduction when $\alpha\geq1$} \label{sec:red_alpha_1}
A similar asymptotic procedure reveals a different reduced system for the $\alpha\geq1$ regime
(see appendix \ref{app:red_alpha_1} for details). 
If $a_n^{(1)}= \eps A_n^{(1)}$ and $\delta = \eps^2 \hat{\delta}$, the  relevant reduced system of nonlinear amplitude equations is now
\begin{equation}
    0=\frac{1}{\hat{\sigma}_n^{(1)}} A_n^{(1)} + \sum_m \NL_{n-m,m}^{111} A_{n-m}^{(1)}A_m^{(1)} +  \langle{\boldsymbol{v}_n^{(1)},\boldsymbol{\hat{f}}}\rangle \delta \delij_{n,\pm 1},
    \label{eq:amplitude_sys}
\end{equation}
where  $\sigma_n^{(1)}=\hat{\sigma}_n^{(1)}/\eps$ and the corresponding  solution is
\begin{equation}
    \boldsymbol{q} =
    \begin{pmatrix}
        v \\
        \eta
    \end{pmatrix}
    = \sum_n \varepsilon A_n^{(1)}  e^{\mathrm{i}n\omega t}\boldsymbol{u}_n^{(1)}+ h.o.t.
    \label{eq:soln_recon_w~lam}
\end{equation}

The important new feature here is that  there is no slow time in this regime as the `fast' timescale already is slow since $\omega \lesssim O(Re^{-1})$; i.e. the forcing period either is comparable to or longer than the timescale for viscous relaxation. 
Consequently, the system can only describe steady amplitude reduced states (\ref{eq:amplitude_sys}) and not the time evolution toward or away from them. Formally, all the  amplitudes $A_n^{(1)}$ need to be included in (\ref{eq:soln_recon_w~lam}) to ensure asymptotic consistency, and system has to be solved numerically to find equilibria (we use the \texttt{fsolve} function in \texttt{MATLAB} with a relative tolerance of $10^{-14}$). Since here the singular values decay algebraically with $n$ (e.g. see figure \ref{fig:sigma_subdom_scaling_PPF}(b)), good agreement with the exact solutions often can be obtained by truncating in $n$, typically with only a few modes ($|n| \leq 2$, for example).

\begin{figure}
    \captionsetup{width=\columnwidth}
    \centering
    \includegraphics[width=1\linewidth]{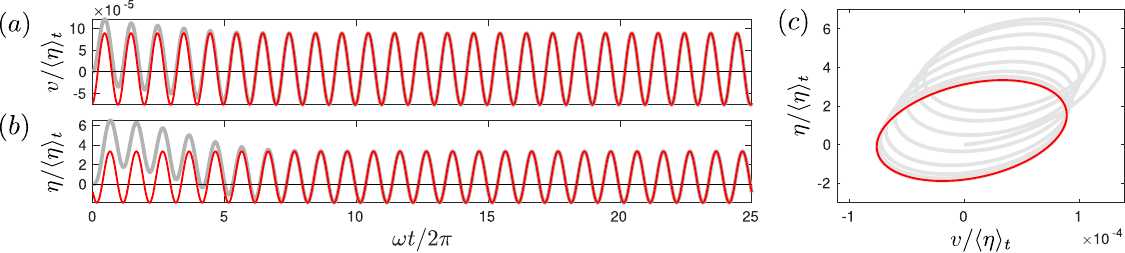}
    \caption{$(a,b)$ Normalised components the solution $\boldsymbol{q}/\langle\eta\rangle_t = (v,\eta)/\langle\eta\rangle_t$ as a function of time for the fully nonlinear solution (grey) and the solution of the amplitude system (Eq. (\ref{eq:amplitude_sys})) (red) for $\Rey=10^{5}$, $\omega = 3/\Rey$, $b=1$, $\hat{\delta} = 5$.  The solid black line shows the applied harmonic forcing as a function of time, which looks close to zero since $\sigma_1^{(1)}=\mathcal{O}(\Rey^2)$ in this case. $(c)$ Phase portrait of the solutions showing the fully nonlinear solution relaxing to the limit cycle captured by the amplitude system.}
    \label{fig:timeseries_w~lam}
\end{figure}

A stable branch of solutions corresponding to a limit cycle in the original $\boldsymbol{q}$ variables exists, just as for $0 < \alpha <  1$, as the forcing $\delta$ is increased from zero (see the solid black line in figure \ref{fig:bif_mean_compare_all}$(b)$). 
A critical maximum forcing amplitude
\beq
\delta_c = \mathcal{O}(\eps^2)= \mathcal{O}(Re^{-4}),
\eeq
set by the saddle node bifurcation where an unstable branch of solutions collides with and annihilates the stable branch, demarcates the end of the weakly nonlinear regime.  
At $\Rey=10^4$ and $\omega = 1/\Rey$, we observe excellent agreement between this critical forcing threshold as predicted by the amplitude equation ($\hat{\delta}_c \approx 1.61937$) and the numerical value obtained from the fully nonlinear system ($\hat{\delta}_c \approx 1.61965$), the solutions of which are compared in figure \ref{fig:bif_mean_compare_all}$(b)$.
As a further test, the evolution of the full nonlinear system (equation (\ref{eq:model-sys})) was computed starting at the origin with $\Rey=10^{5}$, $\omega = 3/\Rey$ ($\alpha=1$), $b=1$, $\hat{\delta} = 5$, yielding the time-series shown in figure \ref{fig:timeseries_w~lam}.  The solution undergoes an initial transient before being attracted to the limit cycle solution of the asymptotically reduced system (equation (\ref{eq:amplitude_sys}))  shown in red.  This result is representative of all such comparisons we performed across a wide range of parameters below the critical forcing threshold; that is, our results indicate that the reduced amplitude equation successfully captures the weakly nonlinear attractor.

%
%

%
%
\subsection{No asymptotic reduction for $\alpha\leq0$} \label{sec:red_alpha<0}

Both leading singular values  scale like  $\Rey^{\alpha}\lesssim \mathcal{O}(1)$ for $\alpha \leq 0$ (equations (\ref{eq:singular_val_scale_cases})) and so $\eps \gtrsim \mathcal{O}(1)$ meaning there is no strong non-normality and therefore no opportunity for asymptotic reduction using the formalism developed herein. 
Nevertheless, since the phase space is available in this simple model system (figure \ref{fig:phase_port_sketch}), we can anticipate that the response will be essentially linear and so $\mathcal{O}(\delta\Rey^{\alpha})$ for small forcing. If, however, the response becomes comparable in magnitude to the basin boundary of $\overline{\bq}_{0}$ i.e. the stable manifold of the saddle point $\overline{\bq}_{-}$, there should be a transition from the weakly nonlinear regime to a fully nonlinear regime where the dynamics converge to $\overline{\bq}_{+}$.  Since $\overline{\bq}_{-} \sim (\Rey^{-3}, \,\Rey^{-2})^T$, the critical forcing amplitude to observe a transition should be $\delta_c = \mathcal{O}(\Rey^{-(\alpha+2)})$.  

\begin{figure}
    \captionsetup{width=\columnwidth}
    \centering
    \includegraphics[width=0.7\linewidth]{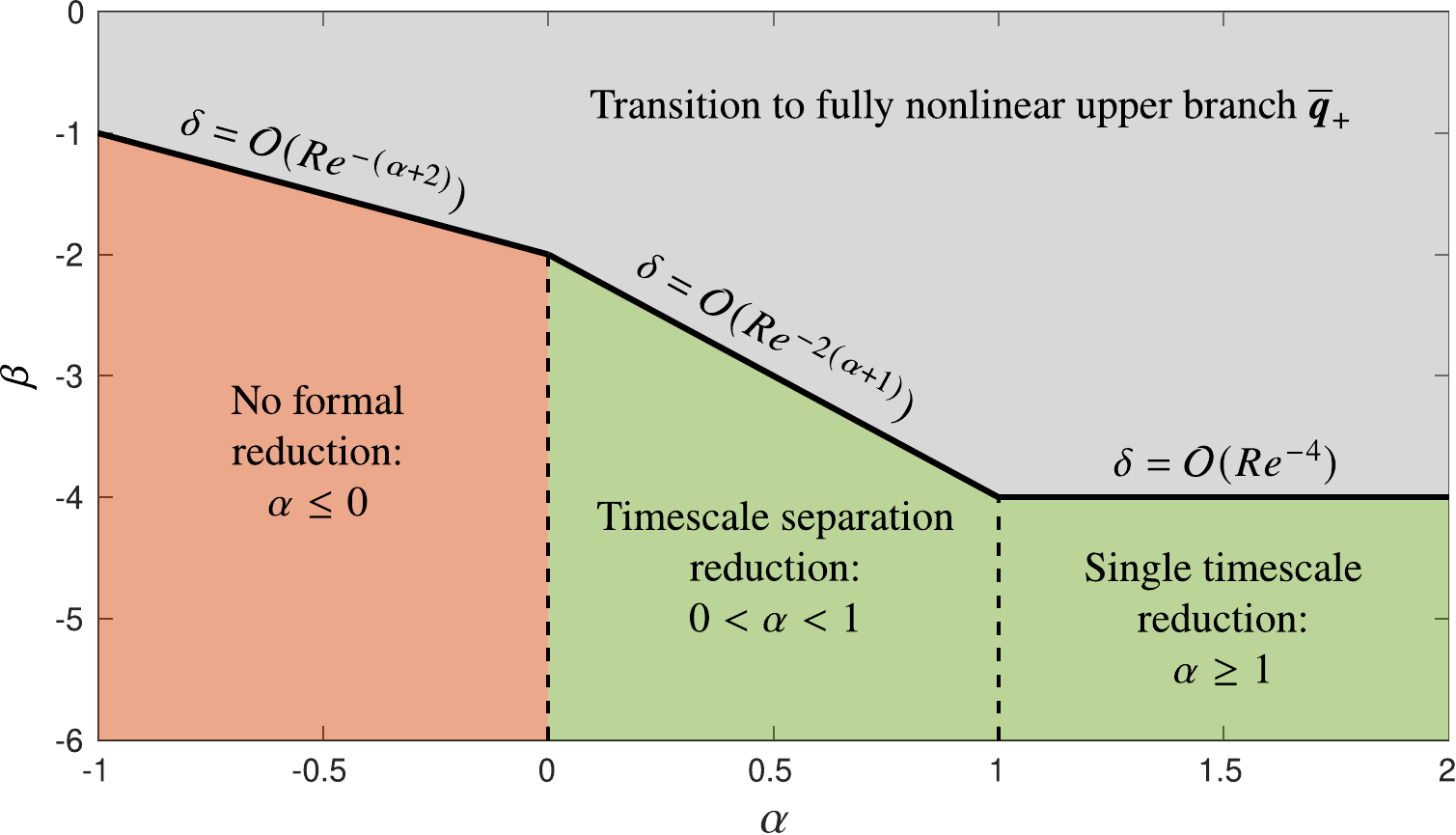}
    \caption{Regime diagram for the behaviour and asymptotic reduction of the system (\ref{eq:model-sys}) as the forcing amplitude $\delta=\mathcal{O}(\Rey^{\,\beta})$ and frequency $\omega = \mathcal{O}(\Rey^{-\alpha})$ are varied.  Green shaded region denotes an asymptotic reduction is possible, red denotes that no formal reduction based on strong non-normality can be achieved, and grey denotes sufficient forcing amplitude that a stable weakly nonlinear solution does not exist, implying a transition to the fully nonlinear upper-branch solution. } 
    \label{fig:regime_map}
\end{figure}

The response of the model system to forcing at any frequency is  summarised in figure \ref{fig:regime_map}.

%
%

%
%
\subsection{Comparison to previous approaches} \label{sec:comparison_model_duc}

The model problem (\ref{eq:model-sys}) can be treated by the approach advocated by \citet{Ducimetiere_PRE_2025}, which leads to the cubic amplitude equation (\ref{eq:duc_amp}).  Asymptotic consistency of this amplitude equation requires that the coefficients $c_0$ and $c_2$ (see equation (\ref{eq:duc_c0_c2})) remain $\mathcal{O}(1)$ as $\varepsilon\rightarrow0$.  Instead, for the model problem, we find that 
\begin{equation}
    c_2 = \mathcal{O}(\varepsilon^{-1}),
\end{equation}
whenever an asymptotic reduction is possible (i.e. $\alpha>0$), and 
\begin{equation}
    c_0 = 
    \begin{cases}
            \mathcal{O}(\varepsilon^{-1/\alpha}), &0< \alpha <1,\\
        \mathcal{O}(\varepsilon^{-1}), & 1 \leq \alpha
    \end{cases}
\end{equation}
as $\varepsilon\rightarrow0$.  Consequently, incorrect terms appear in the leading-order equations of \citet{Ducimetiere_PRE_2025} in both regimes compared to the asymptotically consistent reduced equations derived here.  Specifically, for $0<\alpha<1$, no nonlinear terms should appear in the leading-order equations at $n$\,=\,1 (see equation (\ref{eq:soln_a1_scalar})), an additional $a_0^{(1)}$ nonlinear self-interaction appears at leading-order in the $n$\,=\,0 equation (\ref{eq:a0_equation_model}), and amplitudes with $|n|\geq2$ are subdominant as $\varepsilon\rightarrow0$.  Similarly, when $\alpha\geq1$, all nonlinear terms involving $a_n^{(1)}$ appear in the leading-order equations (\ref{eq:amplitude_sys}).

%
%
\section{Conclusions} \label{sec:conclusion}
In this work, we have developed an asymptotically-consistent, weakly nonlinear analysis (WNA)  for stable but highly non-normal systems subjected to small-amplitude harmonic forcing. The approach is based upon a small parameter $\eps$ defined as the reciprocal of the leading singular value of the resolvent operator corresponding to the forcing frequency.  Contrary to classical WNA, the resulting amplitude equations do not have a universal form, but rather, are system-dependent, and in some cases even vary with the forcing frequency regime for a given system (e.g. (\ref{eq:amp_eqn_scalar}) and (\ref{eq:amplitude_sys}) for the model problem\,). For all the cases considered here, this `non-normal' WNA involves leading responses at frequencies not directly forced which can even dominate those at the forcing frequency (e.g. see (\ref{eq:PPF_solution})\,). This is in stark contrast with `classical' WNA, for which the sole leading amplitude is that of the directly-relevant neutral eigenfunction at the nearby bifurcation point.

Although structurally different reduced equations are obtained for channel flow (see (\ref{eq:PPF_reduced_sys})\,) and a 2-ODE model for two different forcing frequency regimes (see (\ref{eq:amp_eqn_scalar}) and (\ref{eq:amplitude_sys})\,), the qualitative behaviours of these reduced systems exhibit certain similarities.  As the forcing is increased, a response of growing magnitude develops from the original unforced state, which either is predominantly steady or shares the  temporal periodicity of the forcing.  As the forcing is increased further, the system may undergo bifurcations, leading to periodic or aperiodic behaviour.  In all cases studied here, however, this sustained state is destroyed at a critical value of the forcing $\delta_c$ through a saddle-node bifurcation, in which it collides with an unstable invariant solution embedded in the boundary of its basin of attraction.  Above this threshold, no stable weakly-nonlinear solution exists and trajectories from all initial conditions grow into a different, fully nonlinear regime, despite the linear stability of the unforced laminar state. Beneath this forcing threshold, the dynamics of the amplitude equations depend on the initial conditions.  Initial conditions close to the laminar state are typically attracted to the sustained state, whereas larger-amplitude initial conditions can cross the basin boundary  transitioning to fully nonlinear dynamics.  In this sense, the reduced amplitude equations capture subcritical transitions driven by forcing and/or varied initial conditions, and identify critical parameters for which no stable weakly nonlinear state exists.

Relative to previous work, we have obtained asymptotically-consistent amplitude equations that are qualitatively different from those argued for in \cite{ducimetiere2022weak} and \cite{Ducimetiere_PRE_2025} by appealing to `classical' WNA ideas. 
The discrepancy  primarily arises because there is a multi-modal response excited by the interaction of the linear non-normal growth and the nonlinearity, which also introduces non-trivial dependencies on $\varepsilon$ into the equations. This complication is ignored in the work of Ducimeti\`{e}re \emph{et al.} and leads to clear inconsistencies: see \S\ref{sec:duc_compare} and \S\ref{sec:comparison_model_duc}. Despite this, \citet{ducimetiere2022weak} reported surprisingly good agreement with DNS for PPF at low $\Rey$.  We find here that this apparent success can be attributed to the fact that their amplitude equation can be recovered rationally at low $\Rey$ by truncating the governing equations under a low-rank assumption based on the numerical magnitude---rather than the asymptotic scaling---of the singular values, and restricting solutions to equilibria.  

The general method proposed in \S\ref{sec:gen_system} provides a systematic and asymptotically consistent framework for deriving reduced equations in systems exhibiting strong non-normal effects.  By explicitly accounting for the complexity arising from the nonlinear excitation of multi-modal non-normal amplification, the method avoids assumptions of universality and instead adapts naturally to the structure of the underlying system.  The non-universality revealed here may appear daunting compared to classical WNA, yet modern and efficient numerical methods that are highly flexible to the governing equations \citep{dedalus_methods,skene2025fast} are now available for  deriving these reduced equations across a wide class of non-normal, nonlinear systems.  As more of these systems are investigated, it will be of interest to determine whether common structural features emerge that simplify the asymptotic reduction or clarify the extent to which non-universality is an inherent characteristic of these systems.

\backsection[Acknowledgements]{The authors gratefully acknowledge the support of the Geophysical Fluid Dynamics summer program (NSF OCE 1829864) at the Woods Hole Oceanographic Institution, where this work was initiated.  In particular, they would like to thank Basile Gallet for useful discussions in the early stages of this work regarding the model problem in \S\ref{sec:model_prb}.  G.P.C. also acknowledges support from the U.S. Department of Energy through award DE-SC0024572.
M.M. acknowledges the award of a David Crighton fellowship and the kind hospitality of the Department of Applied Mathematics and Theoretical Physics at the University of Cambridge.} 

\backsection[Declaration of interests]{The authors report no conflict of interest.}

%
%

%
%
\appendix

\section{Local scaling in the pre-asymptotic regime} \label{app:local_scaling}

In \S\ref{sec:PPF_pre_asy}, we develop a truncated system of equations that capture the dynamics of the system at low $\Rey$.  However, it is tempting to ask whether a similar set of equations may be obtained using an asymptotic approach, similar to that employed in \S\ref{sec:nonlinear_red_PPF}, but using the \emph{local} scaling of the various terms rather than their asymptotic scaling as $\Rey\rightarrow\infty$.  

In the pre-asymptotic regime, we observed that $\sigma_1^{(1)} = \mathcal{O}(\Rey^{3/2})$ with $\omega = \mathcal{O}(\Rey^{-1/5})$ when $\Rey\lesssim 4\times10^5$.  Then, defining 
\begin{equation}
    \varepsilon \equiv \Rey^{-3/2},
\end{equation}
we numerically assess the scaling of the various singular values with $\varepsilon$, finding locally for $2700\leq\Rey\leq3200$ that
\begin{subequations}
\label{eq:PPF_local_scaling}
\begin{alignat}{1}
    \sigma_1^{(j)} &=
    \begin{cases}
        \mathcal{O}(\Rey^{3/2}) = \mathcal{O}(\varepsilon^{-1}), & j=1, \\
        \mathcal{O}(\Rey^{2/3}) = \mathcal{O}(\varepsilon^{-4/9}), & j=2, \\
        \mathcal{O}(\Rey^{1/3}) = \mathcal{O}(\varepsilon^{-2/9}), & j\geq3, 
    \end{cases}, \\
    \sigma_n^{(j)} &=
    \begin{cases}
        \mathcal{O}(\Rey^{1/2}) = \mathcal{O}(\varepsilon^{-1/3}), & j=\{1,2\}, \\
        \mathcal{O}(\Rey^{1/3}) = \mathcal{O}(\varepsilon^{-2/9}), & j\geq3, 
    \end{cases} 
    \qquad (|n|\geq2).
\end{alignat}
\end{subequations}
The singular values at $n=0$ remain the same as before (equation (\ref{eq:mean_singular_values})), and thus, $\sigma_0^{(j)} = \mathcal{O}(\Rey) = \mathcal{O}(\varepsilon^{-2/3})$.  Therefore, according to these scalings, the largest amplification will be due to $\sigma_1^{(1)}$, followed by $\sigma_0^{(j)}$, and then $\sigma_1^{(2)}$ as $\Rey$ becomes large.  
However, $\Rey$ is not large enough in this regime for the asymptotic scaling to correctly order the magnitude of the singular values, and we observe that the prefactors still have a strong influence.  In fact at $\Rey=3000$, $\sigma_1^{(1)}$ is smaller than $\sigma_0^{(1)}$ ($\sigma_0^{(1)} \approx 3\,\sigma_1^{(1)}$) despite having a larger scaling exponent with $\Rey$.  

Furthermore, we do not see a clearly defined scaling in the nonlinear terms for this range of $\Rey$.  For example, the nonlinear term which couples the $a_1^{(1)}$ and $a_0^{(j)}$ amplitudes in the mean ($n=0$) equations is modulated by the $\tilde{\mu}_{1,-1}^{j11}$ coefficient, which exhibits scalings between $\Rey^{-1}$ and $\Rey$ for the leading five modes ($j\in M_e$) when $2700\leq\Rey\leq4500$, despite the coefficients themselves having similar magnitudes.  Therefore, it appears that it is not possible to use local asymptotic scalings at low $\Rey$ to rationally reduce the system in this regime using any asymptotic approach.

%
%
\section{Singular value decomposition for the model ODE problem} \label{app:svd_model}

We begin by constructing the resolvent operator 
\beq
\mathcal{R}(\mathrm{i}n\omega) := (\mathrm{i}n\omega I - \mathcal{L})^{-1} =\left( \begin{array}{cc}
\alpha_n & 0 \\
\alpha_n \beta_n & \beta_n
\end{array}\right)
\eeq
which has the two eigenvalues $\alpha_n := (\mathrm{i}n\omega+\Rey^{-1})^{-1}$ and $\beta_n := (\mathrm{i}n\omega+b\Rey^{-1})^{-1}$. The square of the singular values are eigenvalues of the Hermitian matrix 
\beq
\mathcal{R}^H\mathcal{R}=
\left(
\begin{array}{cc}
|\alpha_n|^2(1+|\beta_{n}|^2) & \alpha_n^*|\beta_n|^2 \\
\alpha_n |\beta_n|^2 & |\beta_n|^2
\end{array}\right)
\eeq
so
\beq
\sigma_n^2=
\frac{|\alpha_n|^2+|\beta_n|^2+|\alpha_n|^2|\beta_n|^2 \pm \sqrt{(\,|\alpha_n|^2+|\beta_n|^2+|\alpha_n|^2|\beta_n|^2\,)^2-4|\alpha_n|^2|\beta_n|^2}}{2}.
\eeq
Since $b=\mathcal{O}(1)$, $|\alpha_n|^2 \sim |\beta_n|^2$ and so these are both large compared to 1 for strong non-normality and then the two singular values simplify to just 
\beq
\sigma_n^{(1)} \approx |\alpha_n||\beta_n| \quad \& \quad
\sigma_n^{(2)} \approx 1.
\eeq
If $\omega=O(\Rey^{-\alpha})$ the various cases give
\beq
\begin{array}{cclcll}
\alpha <0: & & \alpha_n, \beta_n \sim \Rey^\alpha \ll 1 & \Rightarrow & \sigma_1^{(1)} \sim \Rey^\alpha \ll 1, & \sigma_1^{(2)} \sim \Rey^{\alpha} \ll 1\\
\alpha =0:  &              & \alpha_n, \beta_n \sim 1  & \Rightarrow               & \sigma_1^{(1)} \sim 1, & \sigma_1^{(2)} \sim 1 \\
0<\alpha< 1: &            & \alpha_n, \beta_n \sim \Rey^\alpha \gg 1 & \Rightarrow & \sigma_1^{(1)} \sim \Rey^{2 \alpha} \gg 1, & \sigma_1^{(2)} \sim 1 \\
1 \leq \alpha: &            & \alpha_n, \beta_n \sim \Rey \gg 1  & \Rightarrow      & \sigma_1^{(1)} \sim \Rey^2\gg 1, & \sigma_1^{(2)} \sim 1 
\end{array}
\eeq
The right singular vectors of $\mathcal{R}(\mathrm{i}n\omega)$ are calculated as the right eigenvectors of $\mathcal{R}^H\mathcal{R}$,
\begin{align}
    \boldsymbol{v}_n^{(1)} &\propto  
    \begin{pmatrix}
        \frac{1}{\alpha_n|\beta_n|^2}(\sigma_n^{(1)\, 2}- |\beta_n|^2) \\
        1
    \end{pmatrix}, \\
    \boldsymbol{v}_n^{(2)} & \propto
    \begin{pmatrix}
        \frac{1}{\alpha_n|\beta_n|^2}(\sigma_n^{(2)\,2} - |\beta_n|^2) \\
        1
    \end{pmatrix},
\end{align}
(recall $|\bv_n^{(j)}|^2=1$ but the normalisation has been suppressed for clarity).
The corresponding left singular vectors can then be written as
\beq
    \boldsymbol{u}_n^{(j)} = \frac{1}{\sigma_1^{(j)}}
    \begin{pmatrix}
        \alpha_n & 0 \\
        \alpha_n\beta_n & \beta_n
    \end{pmatrix}
    \boldsymbol{v}_n^{(j)}  \qquad j \in \{1,2 \}.
\eeq
Note, in the regime $0 < \alpha <1$ for example,
\begin{align}
    \boldsymbol{u}_n^{(1)} \propto  
    \begin{pmatrix}
        \mathcal{O}(Re^{-\alpha})\\
        1
    \end{pmatrix}, \quad &
     \boldsymbol{u}_n^{(2)} \propto  
    \begin{pmatrix}
        1\\
        \mathcal{O}(Re^{-\alpha})
    \end{pmatrix}   
    \\
    \boldsymbol{v}_n^{(1)} \propto  
    \begin{pmatrix}
        1 \\
        \mathcal{O}(Re^{-\alpha})
    \end{pmatrix}, \quad &     
    \boldsymbol{v}_n^{(2)} \propto  
    \begin{pmatrix}
        \mathcal{O}(Re^{-\alpha})\\
        1
    \end{pmatrix}
\end{align}
giving 
$\langle \bv_n^{(1)}, \bu_n^{(1)} \rangle =\mathcal{O}(\Rey^{-\alpha})=\mathcal{O}(\sqrt{\eps})$ and $\langle \bv_n^{(1)}, \bu_n^{(2)} \rangle =\mathcal{O}(1)$.

%
%
\section{Left and right singular vector overlap \label{Gelfand}}

Consider a matrix $\A$ which has a dominant singular value $\sigma_1$ so that $\A\approx \sigma_1 \bu_1 \bv_1^H$. Gelfand's formula states that the spectral radius of $\A$ is
\beq
\rho(\A) =\lim_{n \rightarrow \infty} \|\A^n \|^{1/n}
\eeq
for any matrix norm. Using the rank-1 approximation to $\A$ 
\begin{align}
\rho(\A) & \approx \lim_{n \rightarrow \infty} \|(\sigma_1 \bu_1 \bv_1^H)^n \|^{1/n}  \nonumber \\
         & \approx \lim_{n \rightarrow \infty} \|(\sigma_1 \bu_1 (\sigma_1 \bv_1^H \bu_1)^{(n-1)}\bv_1^H \|^{1/n}  \nonumber \\
         & \approx \lim_{n \rightarrow \infty} \sigma_1^{(1-1/n)} |\bv_1^H \bu_1|^{(1-1/n)}\| \A \|^{1/n}  \nonumber \\        
         & \approx \sigma_1 |\bv_1^H \bu_1|
\end{align}
and so
\beq
|\bv_1^H  \bu_1| \approx \frac{\rho(\A)}{\sigma_1}.
\eeq
In terms of the projections for $0 < \alpha < 1$ where $\omega \sim \Rey^{-\alpha}$ and $\eps=\Rey^{-2\alpha}$,
\beq
\begin{array}{lll}
\rho(\,\R(i \omega)\,) \sim \Rey^{\alpha},    &  \sigma_1^{(1)} \sim \Rey^{2 \alpha} & \Rightarrow \quad \rho(\A)/\sigma_1 \sim \Rey^{-\alpha} =\eps^{1/2}\\
\rho(\,\R(0)\,)        \sim \Rey,             & \sigma_0^{(1)} \sim \Rey^2           & \Rightarrow \quad \rho(\A)/\sigma_1 \sim \Rey^{-1}=\eps^{1/2\alpha} 
\end{array}
\eeq
which are the orders of magnitude of $\langle \bv_1^{(1)}, \bu_1^{(1)}\rangle$ and $\langle \bv_0^{(1)}, \bu_0^{(1)}\rangle$ respectively as in (\ref{overlap_11}).

%
%
\section{Asymptotic reduction when $\boldsymbol{\alpha\geq1}$ for the model problem} \label{app:red_alpha_1}

\subsection{Reduction when $\alpha=1$}

When the forcing frequency $\omega =\mathcal{O}(1/\Rey)$ or $\alpha=1$, the leading singular values for each $n$ have the same scaling  
\begin{equation}
    \sigma_n^{(1)} = \mathcal{O}(\eps^{-1}).
\end{equation}
and
\begin{equation}
               \Ov_n^{11} = \mathcal{O}(\varepsilon^{1/2}), 
    \quad\quad \Ov_n^{12} = \mathcal{O}(1).
\end{equation}
Despite these scalings, $a_n^{(2)}$ is seen from the linear analysis to be of size $\varepsilon$ smaller than the leading coefficient $a_n^{(1)}$ so again the dominant time derivatives will involve $a_n^{(1)}$ only. Evaluation of nonlinear interactions of frequencies $n,m \in \mathbb{Z}$ gives  the following scalings
\beq
\NL_{n,m}^{111} = \mathcal{O}(1), \quad
\NL_{n,m}^{121} = \NL_{n,m}^{112} = \mathcal{O}(\varepsilon^{1/2}) \quad \& \quad
\NL_{n,m}^{122} =  \mathcal{O}(\varepsilon).
\eeq
Consequently, the nonlinear interactions between the leading singular vectors dominate in equation (\ref{eq:final_gen_eq}) and so the the leading equations become 
\begin{subequations}
\begin{alignat}{2}
    &|n|=1: \qquad && (\ldots) \Delta \partial_T a_n^{(1)} \varepsilon^{1/2} + (\ldots) \varepsilon a_n^{(1)} - O \big( \sum_m a_{n-m}^{(1)}a_m^{(1)} \big) = \mathcal{O}(\delta ), \qquad \label{eq:balance_n1_a1}\\
    &|n|\neq 1: \qquad && (\ldots) \Delta \partial_T a_n^{(1)} \varepsilon^{1/2} + (\ldots) \varepsilon a_n^{(1)}  = \mathcal{O}\big(\sum_m a_{n-m}^{(1)}a_m^{(1)}\big) \qquad \label{eq:balance_n2_a1}
\end{alignat}
\end{subequations}
with $a_n^{(2)}$ noticeably absent. In the $n=1$ equation, an asymptotic balance between the linear term $\varepsilon a_1^{(1)}$ and the forcing $\delta$ sets $a_1^{(1)} \sim \delta/\varepsilon$ and a nonlinear interaction between the mean $a_0^{(1)}$ and $a_1^{(1)}$ sets the amplitude of $a_0^{(1)}$ as $\mathcal{O}(\varepsilon)$.  In the $n=0$ equation, balancing nonlinearity with the linear term, $a_1^{(1)}a_{-1}^{(1)} \sim (\delta/\varepsilon)^2\sim \varepsilon^2$, gives $\delta\sim\varepsilon^2$, and thus, $a_1^{(1)}\sim\varepsilon$.  Then $\Delta$ must be set as $\Delta\sim\varepsilon^{1/2}$, but $\omega\sim\Rey^{-1}\sim\varepsilon^{1/2}$ and so  there is, in fact, no timescale separation.  This conclusion is consistent with the analysis of the linearised system in \S\ref{sec:model_prb_def}, which showed that the relaxation to the equilibrium state occurs on a $\mathcal{O}(\Rey^{-1})$ timescale; i.e an even slower timescale is not  expected. Finally, in the $n =2 $ equation, the nonlinear interaction between $a_1^{(1)}a_1^{(1)}$ will drive $a_2^{(1)} \sim \eps$ and then, inductively, $a_n^{(1)} \sim \eps$ for all $n$ since
\begin{equation}
      \varepsilon a_n^{(1)} = \mathcal{O}\big(a_{n-1}^{(1)}a_1^{(1)}\big). 
\end{equation}

In conclusion, the only asymptotically consistent choice for scaling of $a_n^{(1)}$, $\Delta$ and $\delta$ for $\omega=\mathcal{O}(\Rey^{-1})$ is:
\begin{my_indent}
    \item All amplitudes have the same asymptotic size $a_n^{(1)} = \mathcal{O}(\varepsilon)$ for $n\in\mathbb{Z}$;
    \item The forcing amplitude depends on $\varepsilon$ as $\delta = \mathcal{O}(\varepsilon^{2})$;
    \item The slow timescale $\Delta =\mathcal{O}(\varepsilon^{1/2}) = \mathcal{O}(\omega)$ which implies that no timescale separation exists.
\end{my_indent}

\subsection{Reduction when $\alpha>1$} \label{sec:app_red_alpha>1}

When $\alpha>1$, the leading singular value  has the same scaling as the previous $\alpha=1$ case so $\sigma_1^{(1)}=\frac{1}{\eps}=\mathcal{O}(\Rey^2)$; see (\ref{eq:singular_val_scale_cases}).  However, additional complexity arises in the scaling of the various inner product terms 
\begin{equation}
    \Ov_n^{11}= \mathcal{O}(\varepsilon^{1/2}+\mathrm{i}\varepsilon^{\alpha/2}), \quad\quad \Ov_n^{12}= \mathcal{O}(1+\mathrm{i}\varepsilon^{(\alpha-1)/2}).
\end{equation}
because the real and imaginary parts of $\alpha_n = (\mathrm{i}n\omega+\Rey^{-1})^{-1}$ and $\beta_n = (\mathrm{i}n\omega+b\Rey^{-1})^{-1}$ now scale differently.
Since, $\alpha>1$, the dominant component of these is the real part of $\Ov_n^{11}a_n$, which exhibits the same scaling as for $\alpha=1$. 
Similar also to the $\alpha=1$ case, the dominant contribution to the nonlinear interactions remains the interaction between the leading left singular vectors such that 
\begin{equation}
    \NL_{n,m}^{111} =  
    \begin{cases}
        \mathcal{O}(1), & n=m=0\\
        \mathcal{O}(1 + \mathrm{i}\varepsilon^{1/2\alpha+1/2}) & n+m= 0,\, n\neq m \\
        \mathcal{O}(1 + \mathrm{i}\varepsilon^{1/2\alpha-1/2}), & \rm{otherwise}
    \end{cases}
    \quad 
\end{equation}
which again maintains the same dominant scaling, i.e. $\NL_{n,m}^{111} = \mathcal{O}(1)$, in its real part.  Thus, for $\alpha>1$, the same asymptotic reduction is obtained as for $\alpha=1$. 

\bibliographystyle{jfm}
\bibliography{jfm}

\end{document}